\documentclass{article}

\usepackage{amsmath,amssymb,amsthm}
\usepackage{graphicx}
\usepackage{caption}
\usepackage{subcaption}
\usepackage{url}
\usepackage{a4wide}

\theoremstyle{definition}

\begin{document}

\title{Coexistence of two dengue virus serotypes\\
and forecasting for Madeira island\thanks{This is a preprint of a paper whose 
final and definite form is in \emph{Operations Research For Health Care}, 
ISSN 2211-6923. Paper submitted 21/Nov/2014; revised 17/May/2015 and 07/Jul/2015;
accepted for publication 14/Jul/2015.}}

\author{Filipa Portugal Rocha$^{1}$\\
{\tt \small filiparocha@meo.pt}
 \and Helena Sofia Rodrigues$^{2, 3}$\\
{\tt \small sofiarodrigues@esce.ipvc.pt}
\and M. Teresa T. Monteiro$^4$\\
{\tt \small tm@dps.uminho.pt}
\and Delfim F. M. Torres$^2$\\
{\tt \small delfim@ua.pt}}


\date{$^1$ \mbox Systems Engineering Master Student, Department of Production and Systems,\\
University of Minho, Campus de Gualtar, 4710--057 Braga, Portugal\\[0.3cm]
$^2${\text{Center for Research and Development in Mathematics and Applications (CIDMA)}},
Department of Mathematics, University of Aveiro, 3810--193 Aveiro, Portugal\\[0.3cm]
$^3$School of Business Studies, Viana do Castelo Polytechnic Institute,\\
Avenida Miguel Dantas, 4930--678 Valen\c{c}a, Portugal\\[0.3cm]
$^4$Algoritmi R\&D Center, Department of Production and Systems,\\
University of Minho, Campus de Gualtar, 4710--057 Braga, Portugal}

\maketitle


\begin{abstract}
The first outbreak of dengue occurred in Madeira Island on 2012,
featuring one virus serotype. Aedes aegypti was the vector
of the disease and it is unlikely that it will be eliminated from the island.
Therefore, a new outbreak of dengue fever can occur and, if it happens,
risk to the population increases if two serotypes coexist. In this paper,
mathematical modeling and numerical simulations are carried out to forecast
what may happen in Madeira Island in such scenario.

\medskip


\smallskip

\end{abstract}


\section{Introduction}

Dengue is the most rapidly spreading mosquito-borne viral disease in the world
\cite{Wearing2006,WHO}. In the last decades, dengue's incidence has grown 30-fold
and is now endemic throughout tropical and subtropical regions,
present in more than 100 countries. According to World Health Organization (WHO),
over 40\% of world's population is at risk \cite{Bhatt2013,Shuman}. In Europe,
in the last decade, some outbreaks have occurred
\cite{Gjenero-Margan2011,Ruche2010,Sofia2014}.

Dengue is transmitted from an infected human to an \emph{Aedes} mosquito,
mainly \emph{Aedes aegypti}, and then from the mosquito on to other humans.
The virus, which is circulating in the blood of the infected subject,
is ingested by a female mosquito when it bites. After the incubation period,
the infectious mosquito can transmit the virus on to other humans
in subsequent bites \cite{Who2009}. The mosquito life cycle has four distinct
stages: egg, larva, pupa and adult. The first three stages take place
in the water, while the adult stage occurs in the air.

Dengue can be caused by four different serotypes, named DEN-1,2,3,4.
It is well documented in the literature that there are differences
among the serotypes (see, e.g., \cite{Balmaseda}), but it is still
to be determined which clinical characteristics correspond to which
serotype \cite{Balmaseda}. However, several reports have indicated
that serotypes DEN-2 and DEN-3 may cause more severe disease than
the serotypes DEN-1 and DEN-4 (see, e.g., \cite{Nisalak,Vaughn}).
Infection by one serotype provides lifelong immunity against that virus and temporary
immunity against the other three. After a cross-immunity period
of 2 to 3 months, a subject can be infected with a different serotype \cite{Feng1997,Watson1999}.
The existence of four different viruses causes a wide clinical spectrum
that includes asymptomatic cases, classic cases of dengue fever and more
severe cases known as Dengue Hemorrhagic Fever (DHF). There is good evidence
that a subsequent infection by other serotype increases the risk of developing
severe dengue \cite{Adams2006,CDC,Esteva2003}.

As there is no vaccine or specific treatment for dengue, prevention is very important.
To prevent the dengue virus transmission, the mosquito vector must be controlled
to avoid reproduction and consequent bites. Control of vector is mainly achieved
by eliminating the places where the mosquito lays their eggs,
such as artificial water containers.
As adult mosquitoes bite inside or outside houses,
during the day or at night when the lights are on,
it is also essential for humans to apply repellent
on skin and also to use personal household protection like window screens,
insecticide treated materials and long-sleeved clothes
\cite{Anderson,Cattand2006,DengueNet}.

Rich literature investigating the dengue disease from the
mathematical point of view exist. Some run simulations on specific regions of the globe
\cite{Harrington2001,Luz2003,Sofia2010c,MyID:168,Sofia2013} and others make
studies using the effects of the disease spread considering biological issues
\cite{Esteva2003,Otero2008,Sofia2009,MyID:263,MyID:283}.
In this paper we aim to study a preview scenario for Madeira Island in the case
of a second outbreak with two serotypes circulating.
This is important because DHF is 15 to 80 times more likely
in secondary infections, being positively associated with pre-existing
dengue virus antibodies \cite{Stephenson2005}.

In the Portuguese island of Madeira, the first outbreak of dengue fever occurred
between October 2012 and February 2013. This outbreak was caused by only one
serotype, DEN-1 virus, and it was declared controlled March 12, 2013.
The vector was the mosquito \emph{Aedes aegypti},
detected in Madeira for the first time in 2005.
The mosquito is still not eradicated from Madeira
and the Portuguese Ministry for Health consider
that it can hardly be \cite{DGS,INSA,carla2012}.
The island has an intensive trade with areas where the disease
is already endemic and the number of tourists has been increasing.
These two factors, beneficial to the local economy, can also
lead to the appearance of a new serotype in the island
and an outbreak can occur at any moment \cite{tese_goncalo}.

The main motivation of this work is to analyze what may happen in Madeira Island
in a dengue outbreak if two serotypes of dengue coexist. Mathematical modeling
is used to simulate this hypothetical outbreak, caused by two different serotypes,
considering the use of insecticide as a control measure.
The article is organized as follows: Section~\ref{sec:2} provides a brief
description of the first outbreak in Madeira and in Section~\ref{sec:3}
we formulate the mathematical model to describe the interaction
between human and vectors. Numerical experiments are presented
in Section~\ref{sec:4} and conclusions ensue in Section~\ref{sec:5}.


\section{Madeira's 2012 outbreak}
\label{sec:2}

\begin{figure}
\center
\includegraphics[scale=0.6]{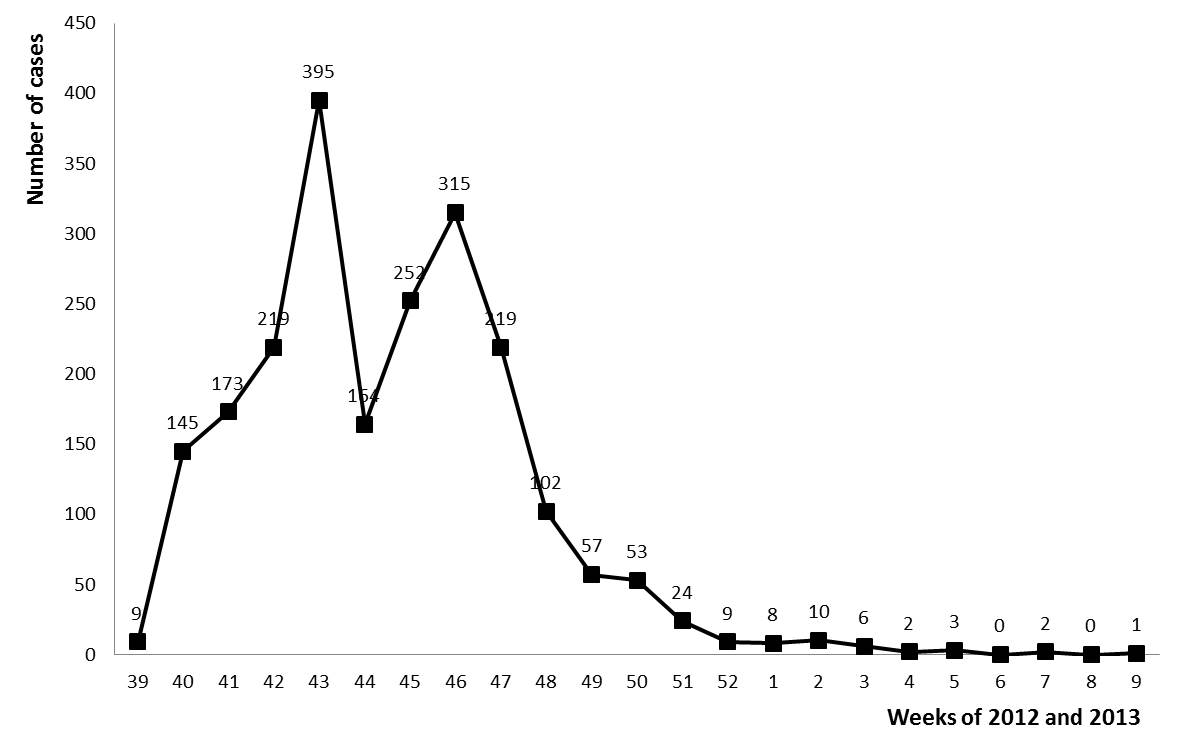}
\caption{Notified dengue fever cases in Madeira per week,
from October 2012 to February 2013 (Source: Instituto de Administra\c{c}\~{a}o da Sa\'{u}de
e Assuntos Sociais, Regi\~{a}o Aut\'{o}noma da Madeira)}
\label{dengue_by_week}
\end{figure}

The outbreak affected mainly Funchal, the main city of the island, and lasted
for 21 weeks, ending late February 2013. By that time, 2171 probable cases
of dengue fever had been reported and, of these, 1084 were confirmed in the laboratory.
In Figure~\ref{dengue_by_week} the reported cases per week are presented.
No deaths or cases of DHF were reported \cite{carla2012,Wilder-Smith}.
In order to control the mosquito vector, local authorities took a whole range of measures,
including the usage of a massive quantity of insecticide, starting even before the outbreak.
However, \emph{Aedes aegypti} has shown a high level of resistance against the
insecticide \cite{tese_goncalo}. As a response, educational campaigns were conducted.
Media coverage in newspapers, television, radio, the internet and through flyers
was a powerful tool. In particular, the population was informed about the
importance of using repellent and covering arms and legs. Artificial water
containers, such as cans, jars, barrels, plastic containers and broken bottles
were requested to be removed. In addition, a medical appointment, exclusively
dedicated to dengue fever, was implemented in \emph{Bom Jesus} Health Centre.
Nowadays, local authorities continue to monitor \emph{Aedes aegypti}
entomological activity through a program of traps monitoring \cite{DGS,carla2012}.


\section{The mathematical model}
\label{sec:3}

In order to describe what can happen if two virus serotypes coexist,
a mathematical model based on \cite{Sofia2014} is here proposed.
The novelty is the presence of a second serotype
of dengue and the consequent presence of a new set of variables:

\begin{tabular}{lcl}
$S_h$ &---& susceptible human (who can contract the disease);\\
$I_{1h}$ &---& human infected by serotype $1$ (who can transmit the serotype $1$);\\
$R_{1h}$ &---& human resistant to serotype $1$ (infected by serotype $1$ and recovered);\\
$I_{jh}$ &---& human infected by serotype $j$ (who can transmit the serotype $j$);\\
$R_{jh}$ &---& human resistant to serotype $j$ (infected by serotype $j$ and recovered);\\
$I_{1jh}$ &---& human infected first by serotype $1$ and after by serotype $j$;\\
$R_{1jh}$ &---& human first resistant to serotype $1$ and after to serotype $j$;\\
$I_{j1h}$ &---& human first infected by serotype $j$ and after by serotype $1$;\\
$R_{j1h}$ &---& human first resistant to serotype $j$ and after to serotype $1$;\\
$S_m$ &---& susceptible mosquitoes;\\
$A_m$ &---&  aquatic phase (stages that take place in water: egg, larva and pupa);\\
$I_{1m}$ &---&  mosquitoes infected by serotype $1$ (that can transmit the disease);\\
$I_{jm}$ &---&  mosquitoes infected by serotype $j$ (that can transmit the disease).\\
\end{tabular}

\medskip

\noindent The index $1$ refers to the virus serotype DEN-1, the responsible for the first outbreak in Madeira Island.
The new index $j$, $j\in \{2,3,4\}$, refers to different virus serotypes (DEN-2, 3 or 4).

The first nine state-variables are related to human population while
the last four are concerned with the female mosquito.
These compartments are mutually-exclusive \cite{Diekmann}.
It should be noted that there is no variable to represent resistant mosquitoes
or to represent mosquitoes infected by a second serotype. Indeed, given that mosquitoes
have a very short lifespan, there is no time for them to recover from an infection.
Thus, once infected, the mosquito remains infected until its death
\cite{Chan2012,Who2009}.

The application of insecticides is the most common control measure.
In order to simplify the model, insecticide was the only control measure
considered. More precisely, we consider $c_m(t)$ to be the proportion
of insecticide applied at time $t$, with $0\leq c_m(t)\leq 1$.

The system of differential equations for the human population is:
\begin{equation}
\label{human}
\begin{tabular}{l}
$
\left\{
\begin{array}{l}
\displaystyle \frac{dS_h}{dt} = \mu_h N_h- \left(B \beta_{1mh} \frac{I_{1m}}{N_h}+B\beta_{jmh}\frac{I_{jm}}{N_h}+\mu_h\right)S_h\\
\displaystyle \frac{dI_{1h}}{dt} = B\beta_{1mh}\frac{I_{1m}}{N_h}S_h - \left(\eta_{1h} + \mu_h\right)I_{1h}\\
\displaystyle \frac{dI_{jh}}{dt} = B \beta_{jmh}\frac{I_{jm}}{N_h}S_h - \left(\eta_{jh} + \mu_h\right)I_{jh}\\
\displaystyle \frac{dR_{1h}}{dt} = \eta_{1h} I_{1h} - \left(\sigma B\beta_{jmh}\frac{I_{jm}}{N_h}+\mu_h\right)R_{1h}\\
\displaystyle \frac{dR_{jh}}{dt} = \eta_{jh}I_{jh} - \left(\sigma B\beta_{1mh}\frac{I_{1m}}{N_h}+\mu_h\right)R_{jh}\\
\displaystyle \frac{dI_{1jh}}{dt} = \sigma B\beta_{jmh}I_{jm}\frac{R_{1h}}{N_h}-\left(\mu_h+\mu_{dhf}+\eta_{jh}
\right)I_{1jh}\\
\displaystyle \frac{dI_{j1h}}{dt} = \sigma B\beta_{1mh}I_{1m}\frac{R_{jh}}{N_h}-\left(\mu_h+\mu_{dhf}+
\eta_{1h}\right)I_{j1h}\\
\displaystyle \frac{dR_{1jh}}{dt} = \eta_{jh}I_{1jh}-\mu_h R_{1jh}\\
\displaystyle \frac{dR_{j1h}}{dt} = \eta_{1h}I_{j1h}-\mu_hR_{j1h}\\
\end{array}
\right.$
\end{tabular}
\end{equation}
while for the vector is given by
\begin{equation}
\label{mosquito}
\begin{tabular}{l}
$
\left\{
\begin{array}{l}
\displaystyle\frac{dA_m}{dt} = \varphi \left(1-\frac{A_m}{k N_h}\right)\left(S_m+I_{1m}+I_{jm}\right)
-(\eta_A+\mu_A) A_m\\
\displaystyle\frac{dS_m}{dt} = \eta_A A_m
-\left(B \frac{\beta_{1hm}\left(I_{1h}+I_{j1h}\right)+\beta_{jhm}\left(I_{jh}+I_{1jh}\right)}{N_h}+\mu_m + c_m\right) S_m\\
\displaystyle\frac{dI_{1m}}{dt} = B \beta_{1hm}\frac{\left(I_{1h}+I_{j1h}\right)}{N_h}S_m
-\left(\mu_m + c_m\right) I_{1m}\\
\displaystyle\frac{dI_{jm}}{dt} = B \beta_{jhm}\frac{\left(I_{jh}+I_{1jh}\right)}{N_h}S_m
-\left(\mu_m + c_m\right) I_{jm}.
\end{array}
\right.$
\end{tabular}
\end{equation}
The differential equations are subject to the initial conditions
\begin{equation*}
\begin{tabular}{l l l l l}
$S_h(0)=S_{h0}$,  & $I_{1h}(0)=I_{1h0}$, & $I_{jh}(0)=I_{jh0}$, & $I_{1jh}(0)=I_{1jh0}$, & $I_{j1h}(0)=I_{j1h0}$,\\
$R_{1h}(0)=R_{1h0}$, & $R_{jh}(0)=R_{jh0}$, & $R_{1jh}(0)=R_{1jh0}$, & $R_{j1h} (0)=R_{j1h0}$, & \\
$A_m(0)=A_{m0}$, & $S_m(0)=S_{m0}$, & $I_{1m}(0)=I_{1m0}$, & $I_{jm}(0)=I_{jm0}$. &\\	
\end{tabular}
\end{equation*}
A wide range of parameters is used to build the model, e.g.,
by $N_h$ we denote the total human population in Madeira; mortality is equal to the inverse
of lifespan with a constant rate $\mu_h$ to human population and $\mu_m$
for the mosquito population; an infected mosquito can transmit dengue to
a susceptible human with a probability $\beta_{mh}$ per bite while an infected human
transmits the disease to a susceptible mosquito with a probability $\beta_{hm}$
per bite. Table~\ref{parameter} presents the model parameters, their meaning
and their values. An explanation is, however, necessary.

There is evidence that a secondary infection
increases the risk of developing DHF, which can be explained by
the ADE phenomenon (\emph{Antibody-Dependent Enhancement}) \cite{Wahala2011}.
Recovering from one serotype of dengue, the immune system responds by producing
antibodies to the virus, gaining lifelong immunity against that serotype.
However, when a new serotype of dengue is contracted, the same antibodies
that protect against the previous serotype, facilitate the entry of the new
virus into host cells, enhancing the infection, with an increase in viral production.
Viral loads are associated with transmissibility, and several researchers believe that
subjects are more infectious in the second infection than during the first,
and consequently the transmission rate increases \cite{Bianco2009}.
In this paper we use $\mu_{dhf}$ to represent death from DHF and $\sigma$ to represent
the ADE phenomenon \cite{Adams2006,Esteva2003}.
Note that the total population is not constant in our model
because we are considering a probability of death from DHF different from zero.
This is in contrast with \cite{Sofia2014}. Our approach to the multistrain model
consists of using different parameter values for different strains:
$\eta_{1h} \ne \eta_{jh}$ and/or $\beta_{1mh} \ne \beta_{jmh}$.
For simplicity of notation, in Table~\ref{parameter} we
omit the specific strain $1$ or $j$, $j \in \{2, 3, 4\}$.
\begin{table}[ptbh]
\begin{center}
\caption{Parameters of the epidemiological model \eqref{human}--\eqref{mosquito}}
\label{parameter}
\begin{tabular}{lllll}
\hline
Para- & Description & Range of values  & Used & Source\\
meter& & in the literature & values & \\ \hline
$N_h $ & total population & & 112000 & \cite{censos2011}\\
$B$ & average daily biting (per day)& & 1/3 & \cite{Focks2000}\\
$\beta_{mh}$ & transmission probability from $I_m$ (per bite)& [0.25, 0.33] & 0.25 & \cite{Focks2000}\\
$\beta_{hm}$ & transmission probability from $I_h$ (per bite)& [0.25, 0.33] & 0.25 & \cite{Focks2000}\\
$1/\mu_{h}$ & average lifespan of humans (in days)& & $79\times365$ & \cite{censos2011} \\
$1/\eta_{h}$ & average viremic period (in days)& [4, 15] & 7 & \cite{Chan2012}\\
$1/\mu_{m}$ & average lifespan of adult mosquitoes (in days)& [8, 45]& 15 & \cite{Focks1993,Harrington2001,Freitas2007}\\
$\varphi$ & number of eggs at each deposit per capita (per day)& & 6 & \cite{Sofia2012}\\
$1/\mu_{A}$ & natural mortality of larvae (per day)& & 0.2363 & \cite{Barrios2013}\\
$\eta_{A}$ & maturation rate from larvae to adult (per day)& [1/11, 1/7]& 1/9 & \cite{Padmanabla2011}\\
$k$ & number of larvae per human & & 0.9 & \cite{Focks1995,Watson1999}\\
$\sigma$ & ADE phenomenon & [0, 5] & 0,5; 1.1; 2.5 & \cite{Nuraini2007}\\
$\mu_{dhf}$ &  probability of death from DHF &  [0.01, 0.1] & $0.02$ & \cite{CDC,Who2009}\\
\hline
\end{tabular}
\end{center}
\end{table}

\begin{figure}
\centering
\includegraphics[scale=0.52]{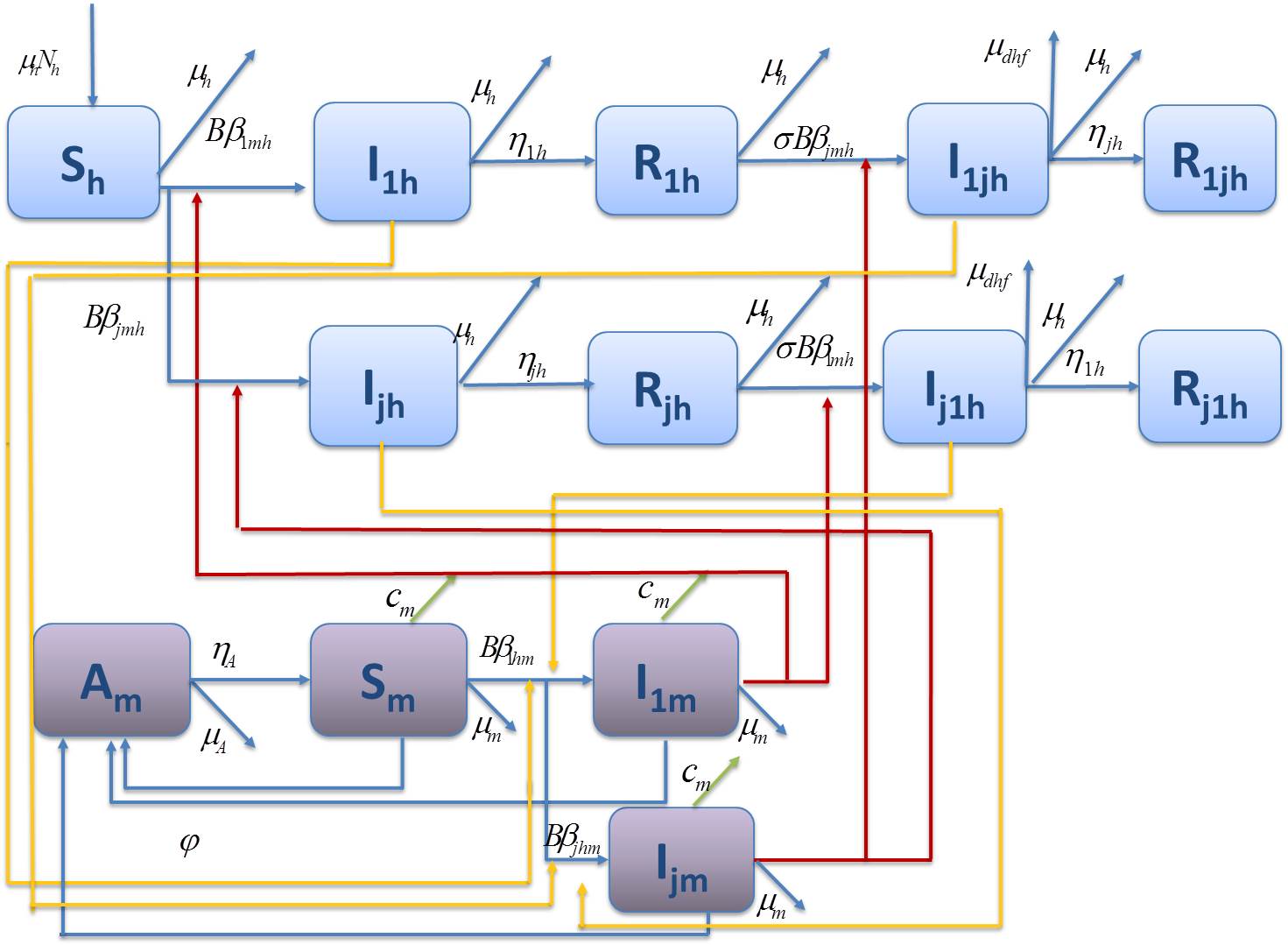}
\caption{The scheme of the dengue model \eqref{human}--\eqref{mosquito} with two serotypes}
\label{coexistence}
\end{figure}

In our model we assume that:
\begin{itemize}
\item Infected mosquitoes do not transmit the virus onto their eggs;
\item The population is homogeneous;
\item There is no immigration or emigration;
\item There is no seasonality;
\item There is an homogeneous mixing of human and mosquito population,
having the vector an equal probability to bite any human;
\item Humans and mosquitoes are assumed to be born susceptible.
\end{itemize}
The scheme of the model \eqref{human}--\eqref{mosquito} is shown in Figure~\ref{coexistence}.
An arrow pointing into a compartment is associated with a positive member of the corresponding
differential equation while an arrow pointing out of the compartment represents a negative member of the equation.

Comparing our model with the previous one \cite{Sofia2014}, this one is more complex.
The coexistence of two serotypes of dengue leads to 13 state variables,
more than doubling the 6 state variables in the previous model \cite{Sofia2014}.
As a consequence, the interaction between human and vector has become more complex
and the number of arrows in Figure~\ref{coexistence} has increased with respect
to the scheme of \cite{Sofia2014}. It is also noted that there are no
equilibrium points and, therefore, the basic reproduction number based
on \cite{Driessche2002} cannot be computed. This is also in contrast with \cite{Sofia2014}.

The impact of ADE on the model can be profound: single strain
models usually cannot have isolated equilibria and show fluctuations without external
seasonal drives or noise, while multistrain models with ADE phenomenon may
present oscillations \cite{Bianco2009}.


\section{Computational experiments}
\label{sec:4}

The software used in our simulations was \textsf{Matlab} with the routine \texttt{ode45}.
This solver is based on an explicit (4,5) Runge--Kutta formula, the Dormand--Prince pair.
The numerical solver \texttt{ode45} combines fourth and fifth order methods, both of which
are similar to the classical fourth-order Runge--Kutta method. These vary the step size,
choosing it at each step in an attempt to achieve the desired accuracy.

\begin{figure}
\centering
\begin{subfigure}[b]{0.45\textwidth}
\centering
\includegraphics[scale=0.41]{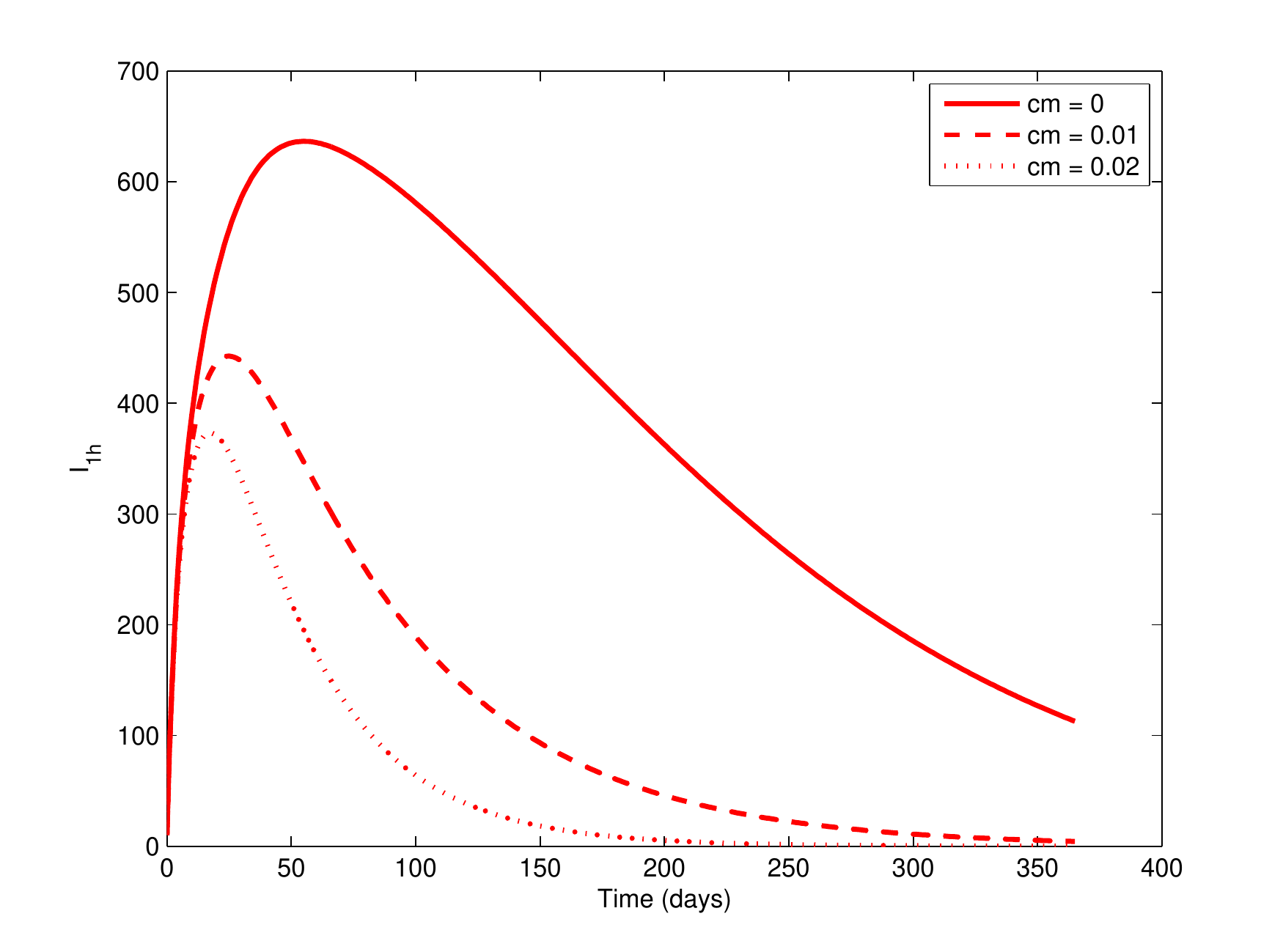}
\caption{$I_{1h}$ (infected by DEN-$1$)}
\end{subfigure}%
\begin{subfigure}[b]{0.45\textwidth}
\centering
\includegraphics[scale=0.41]{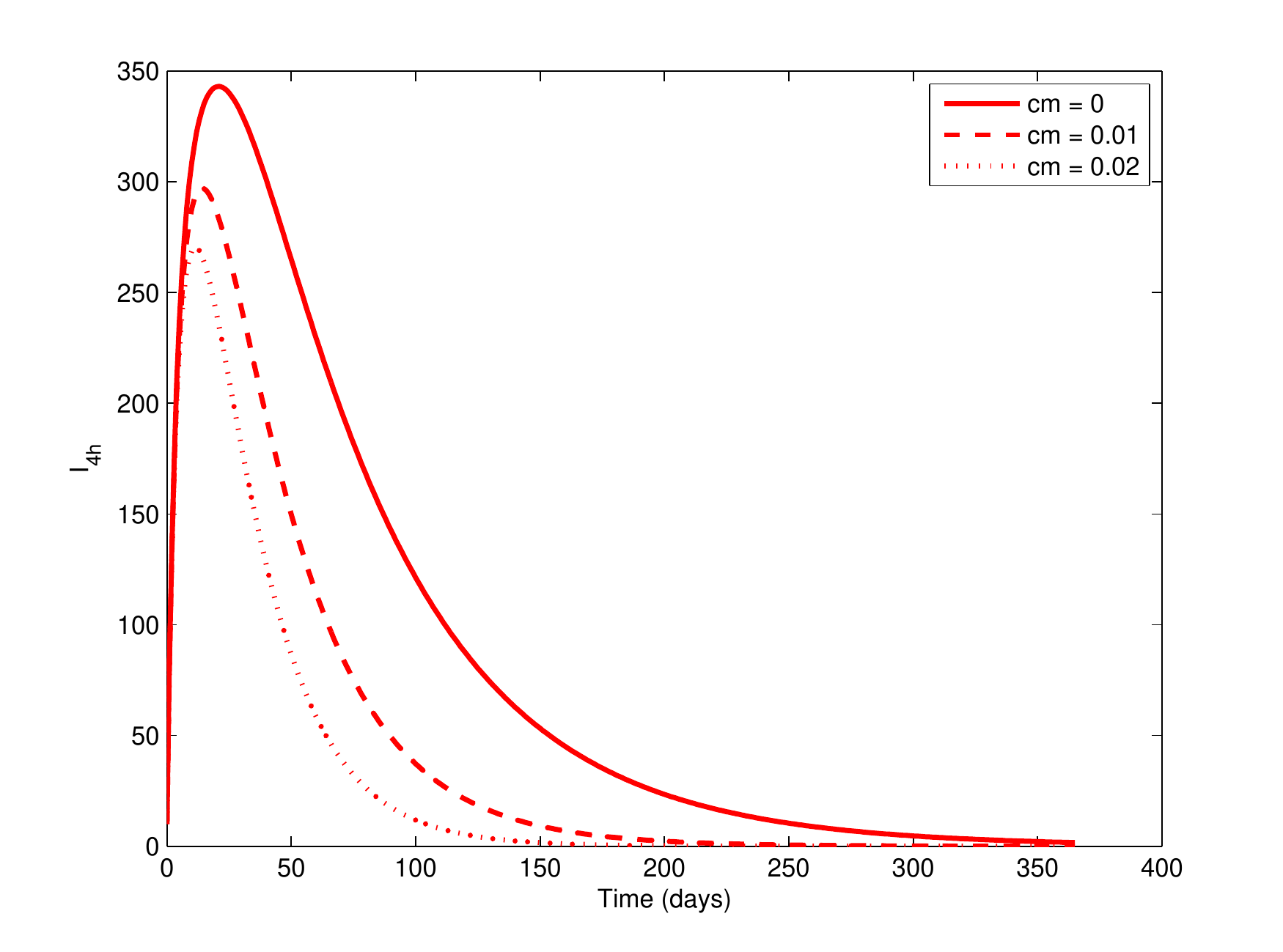}
\caption{$I_{jh}$ (infected by DEN-$4$)}
\end{subfigure}\\
\begin{subfigure}[b]{0.45\textwidth}
\centering
\includegraphics[scale=0.41]{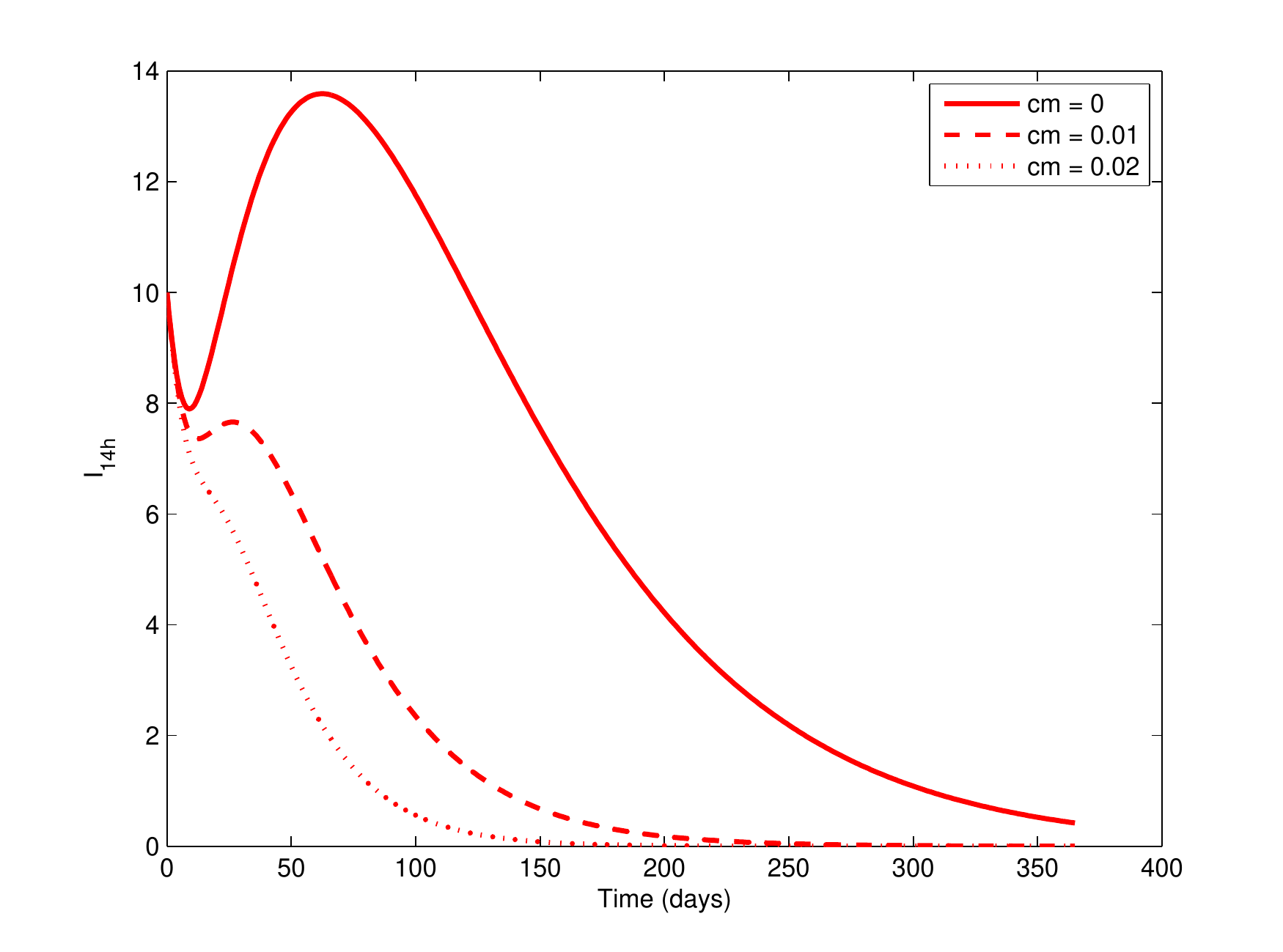}
\caption{$I_{1jh}$ (infected by DEN-$1$, then by DEN-$4$)}
\end{subfigure}
\begin{subfigure}[b]{0.45\textwidth}
\centering
\includegraphics[scale=0.41]{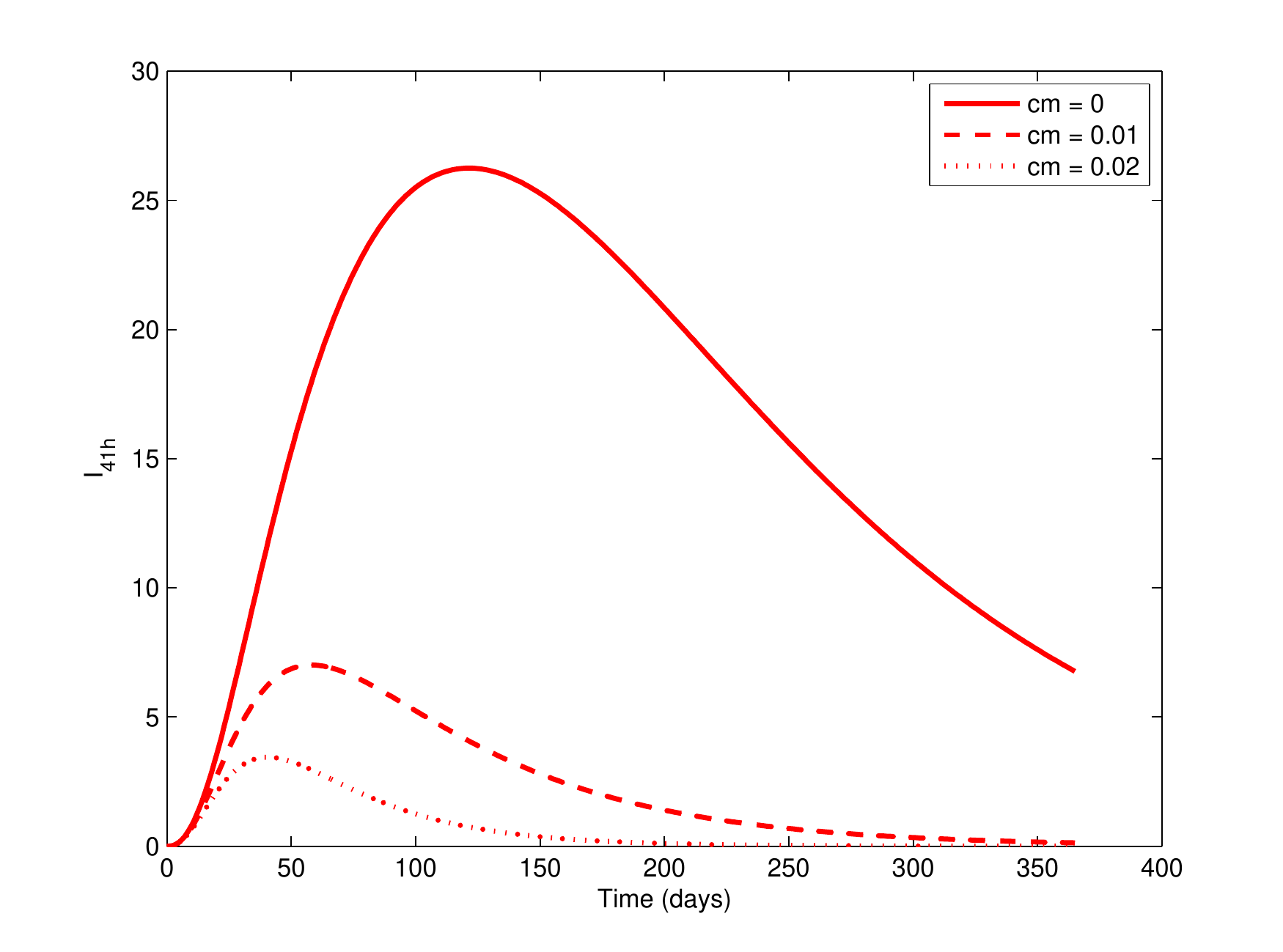}
\caption{$I_{j1h}$ (infected by DEN-$4$, then by DEN-$1$)}
\end{subfigure}
\caption{Infected individuals varying the control $c_m$
in case of two dengue serotypes: DEN-$1$ and DEN-$4$
(results for 1 year with
$\mu_{dhf}=0.02$,
$\sigma=1.1$,
$\beta_{1mh}=\beta_{1hm}=0.25$,
$\beta_{jmh}=\beta_{jhm}=0.25$,
$\eta_{1h}=1/7$,
$\eta_{jh}=1/5$,
and $I_{1h0} = I_{jh0} = I_{1jh0} = 10$)}
\label{var_cmbeta025sigma11}
\end{figure}

Our computational experiments simulate a hypothetical outbreak in Madeira Island
caused by two different serotypes 1 and $j$, $j \in \{2, 3, 4\}$. We considered
that the initial values related with serotype DEN-1 are the final values
of Madeira's outbreak of 2012. Funchal has a total population of 112 000.
Of these, 2171 were infected. Serotype DEN-$j$ is new, so $I_{j1h}(0)=0$.
As the other initial values are unknown, we presumed the following values:
\begin{itemize}
\item $I_{1h}(0)=\xi$ and $I_{jh}(0)=\xi$, $\xi \in \{10, 50, 100\}$
(hypothetical values, assuming that with these infected numbers,
health authorities could act and take measures to control the disease);  	

\item $I_{1jh}(0)=\xi$, $\xi \in \{10, 50, 100\}$
(individuals who have recovered from the outbreak and are now infected by serotype $j$).
\end{itemize}
As the first outbreak caused no deaths, all the infected have recovered from serotype $1$.
However, $\xi$ of these are now infected by serotype $j$, so
\[
R_{1h}(0)= 2171 - \xi,	
\quad R_{jh}(0)= 0,
\quad R_{1jh}(0)= 0,	
\quad R_{j1h}(0)= 0.
\]
Thus, the initial value for susceptible humans is $S_h(0)= 109829 -2 \xi$.
We considered different virulence and transmission intensities.
Two scenarios are possible: or $j \in \{2, 3\}$, and $1/\eta_{jh}$ and $\beta_{jmh}$
are not lower than those of DEN-1 (more aggressive case); or $j = 4$, and $1/\eta_{jh}$
and $\beta_{jmh}$ do not have higher values than those of DEN-1 (less aggressive case).
In order to evaluate the influence of the $\beta$ and $\sigma$ parameters
and the insecticide control $c_m$, some numerical tests were carried out.
At a first step, we fixed $\beta=0.25$ and $\sigma=1.1$ and tested the model
varying the control $c_m$. The corresponding results
are shown in Figure~\ref{var_cmbeta025sigma11}.
It should be noted that in Figures~\ref{var_cmbeta025sigma11},
\ref{var_cmbeta033sigma11}, \ref{var_sigmabeta025cm001} and
\ref{var_sigmabeta033cm001}, the scale of the infected human axis is
smaller in secondary infections than in first infections. This was
essential due to the following reasons:
\begin{itemize}
\item in the first infection people are naive to the disease;

\item there are a substantial number of resistant individuals;

\item there is a partial and transient immunity from the first infection
to the second one, which delays second infections.
\end{itemize}

Figure~\ref{var_cmbeta025sigma11}
shows the fluctuation of the number of infected human
with the application of different proportions of insecticide.
It is patent that insecticide is a very efficient measure.
Even with small quantities, its influence is huge.
Note that with $c_m=0.02$, in the considered time interval,
the number of infected is tending to zero.

It is clear that should a new outbreak of dengue with two different serotypes
occur, health authorities should pay particular attention to the more aggressive serotype:
if the new serotype is DEN-2/3, then the priority is to take care
of people infected by DEN-2/3 (see Figure~\ref{var_cmbeta033sigma11});
if the new serotype is DEN-$4$, then health authorities should pay particular
attention to DEN-$1$ (see Figure~\ref{var_cmbeta025sigma11}).
In all cases, the control has an important role on the number of infected
people and ``doing nothing'' should not be an option to health authorities.

\begin{figure}
\centering
\begin{subfigure}[b]{0.45\textwidth}
\centering
\includegraphics[scale=0.41]{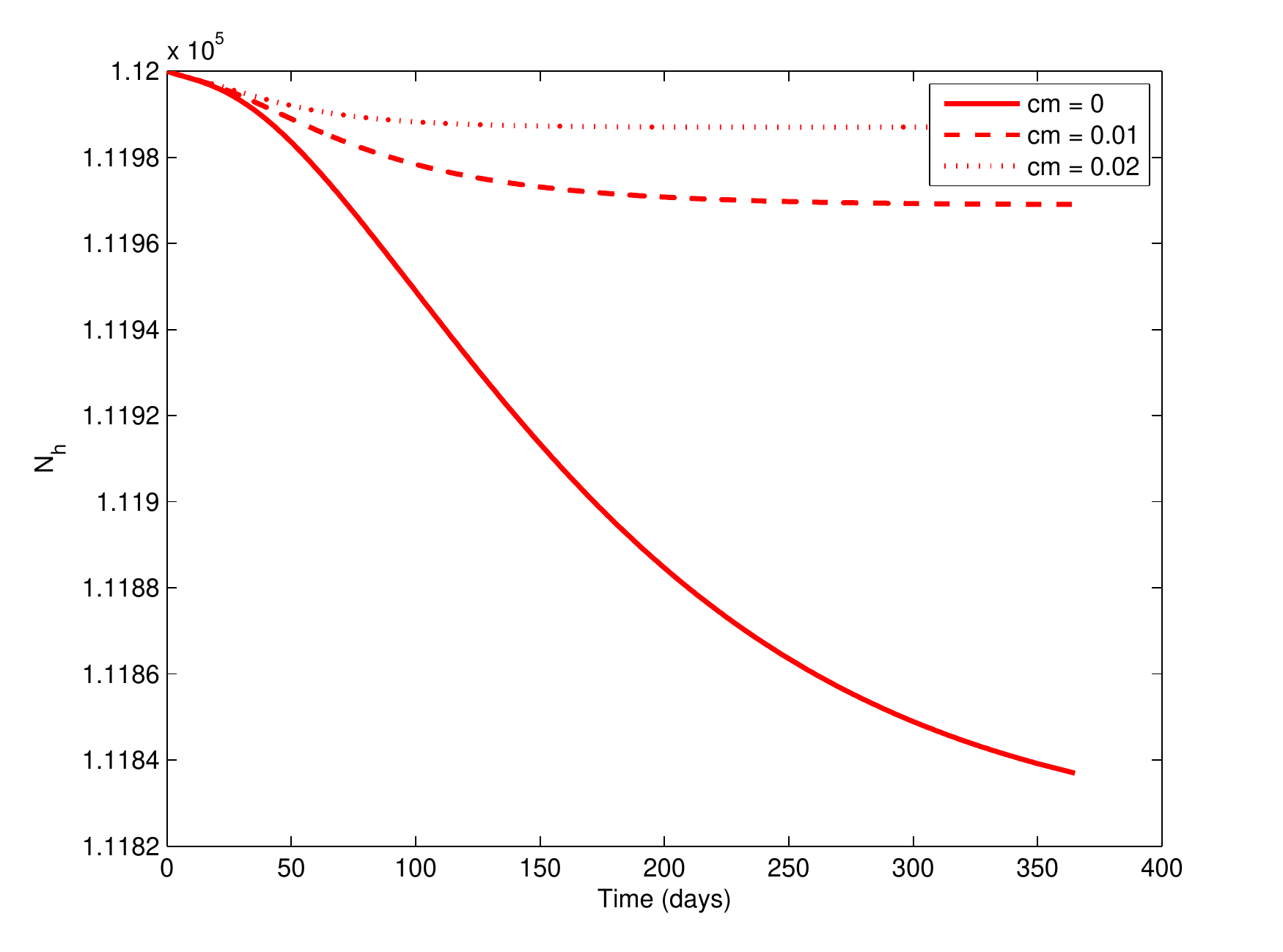}
\caption{DEN-$1$ and DEN-$4$\newline
($\beta_{jmh}=\beta_{jhm}=0.25$, $\eta_{jh}=1/5$)}
\label{var_nh1}
\end{subfigure}%
\begin{subfigure}[b]{0.45\textwidth}
\centering
\includegraphics[scale=0.49]{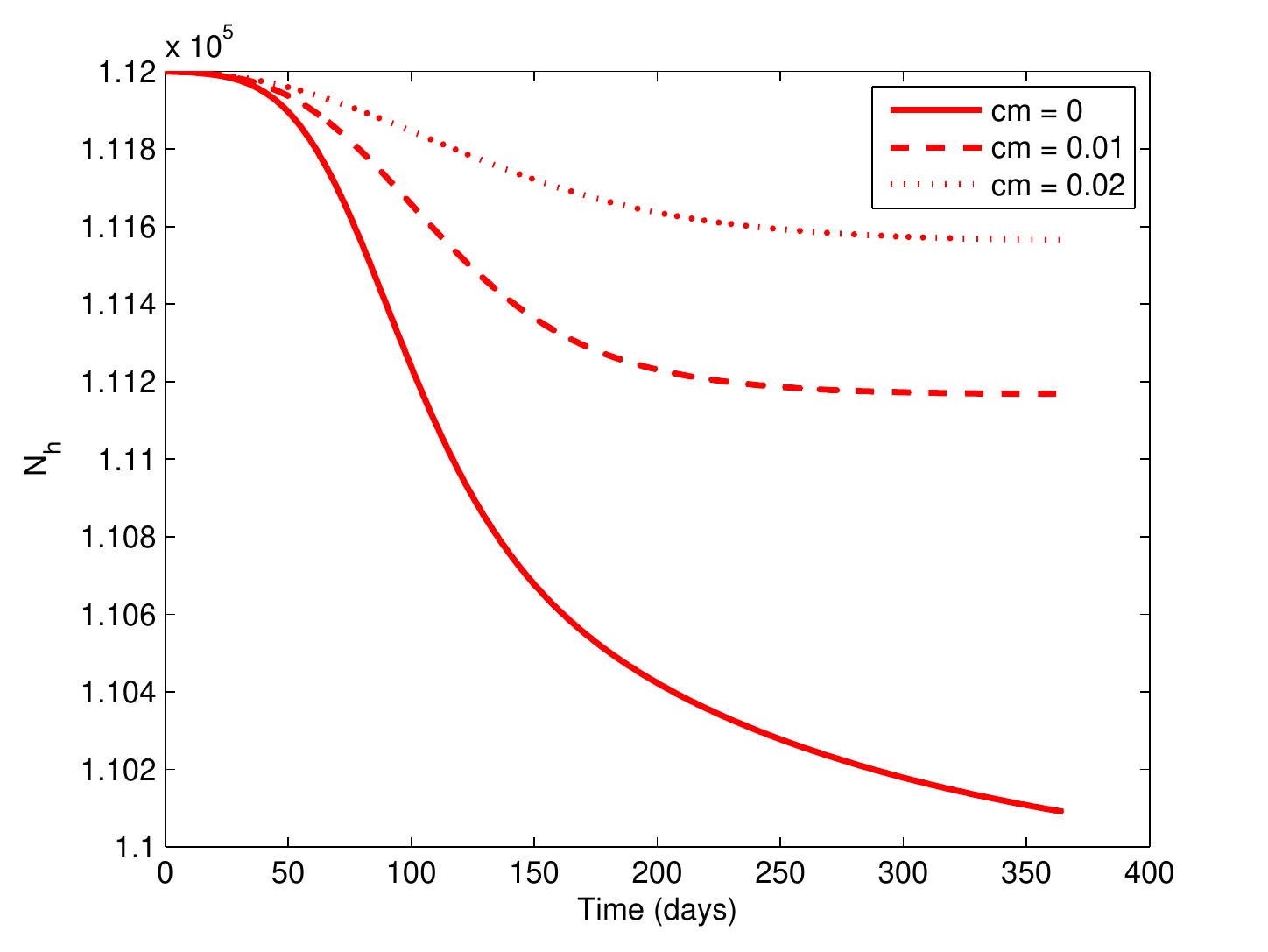}
\caption{DEN-$1$ and DEN-$j$, $j\in \{2, 3\}$\newline
($\beta_{jmh}=\beta_{jhm}=0.33$,
$\eta_{jh}=1/9$)}
\label{var_nh2}
\end{subfigure}
\caption{Total population $N_{h}$ varying the control $c_m$
(results for 1 year with
$\mu_{dhf}=0.02$,
$\sigma=1.1$,
$\beta_{1mh}=\beta_{1hm}=0.25$,
$\eta_{1h}=1/7$,
$I_{1h}(0) = I_{jh}(0) = I_{1jh}(0) = 10$)}
\label{fig46}
\end{figure}

\begin{figure}
\centering
\begin{subfigure}[b]{0.45\textwidth}
\centering
\includegraphics[scale=0.41]{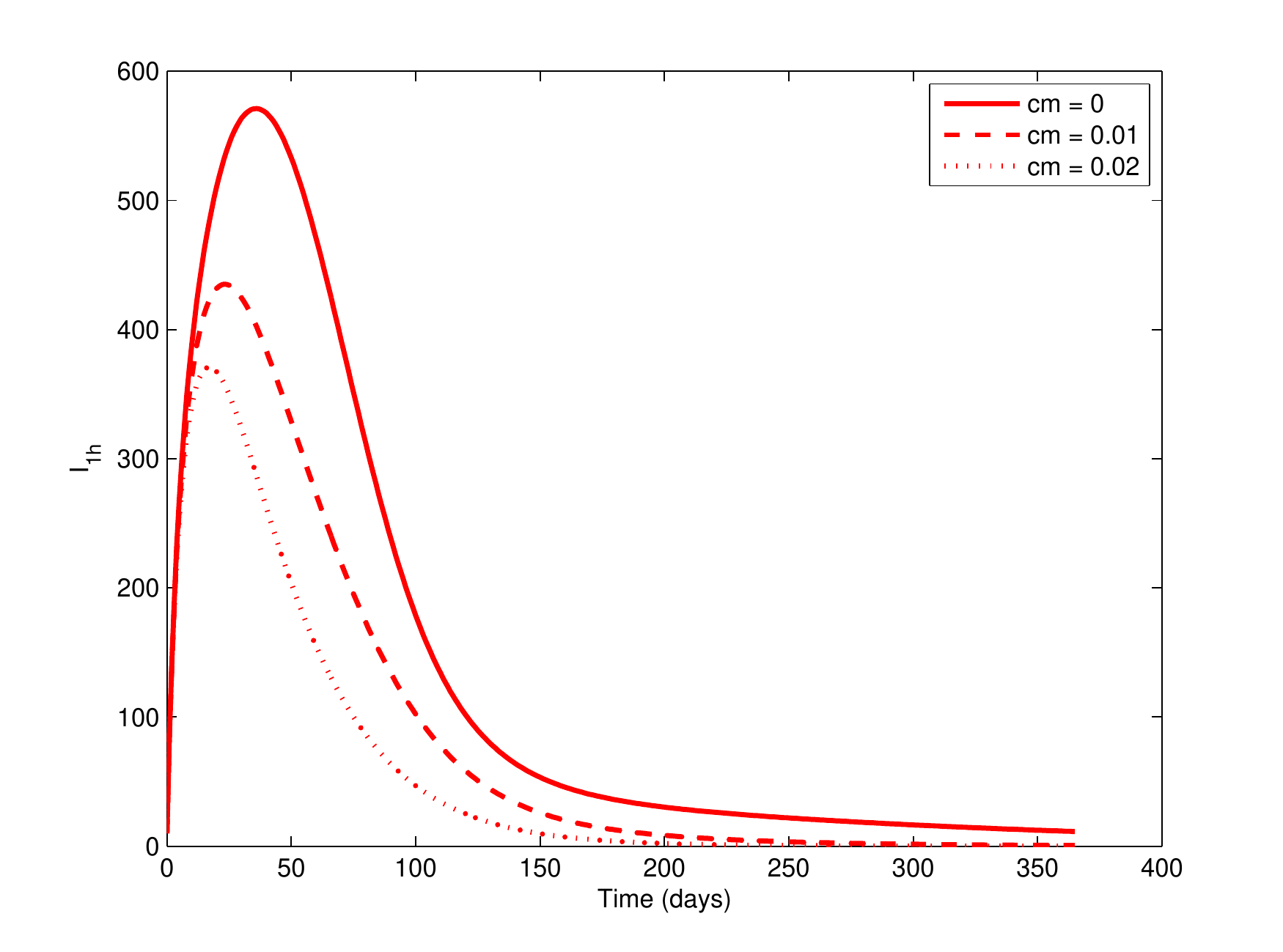}
\caption{$I_{1h}$ (infected by DEN-$1$)}
\end{subfigure}%
\begin{subfigure}[b]{0.45\textwidth}
\centering
\includegraphics[scale=0.41]{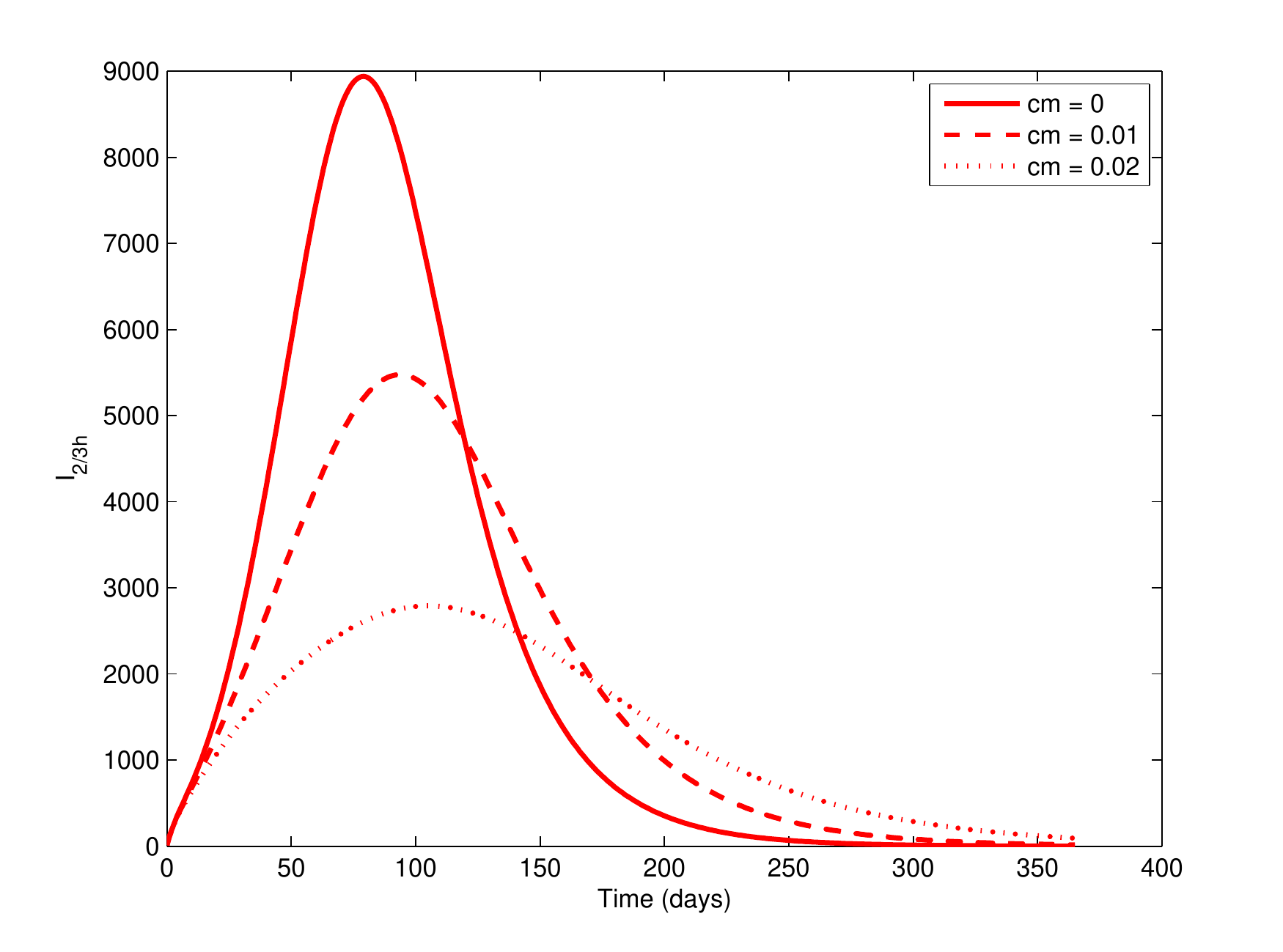}
\caption{$I_{jh}$ (infected by DEN-$j$)}
\end{subfigure}\\
\begin{subfigure}[b]{0.45\textwidth}
\centering
\includegraphics[scale=0.41]{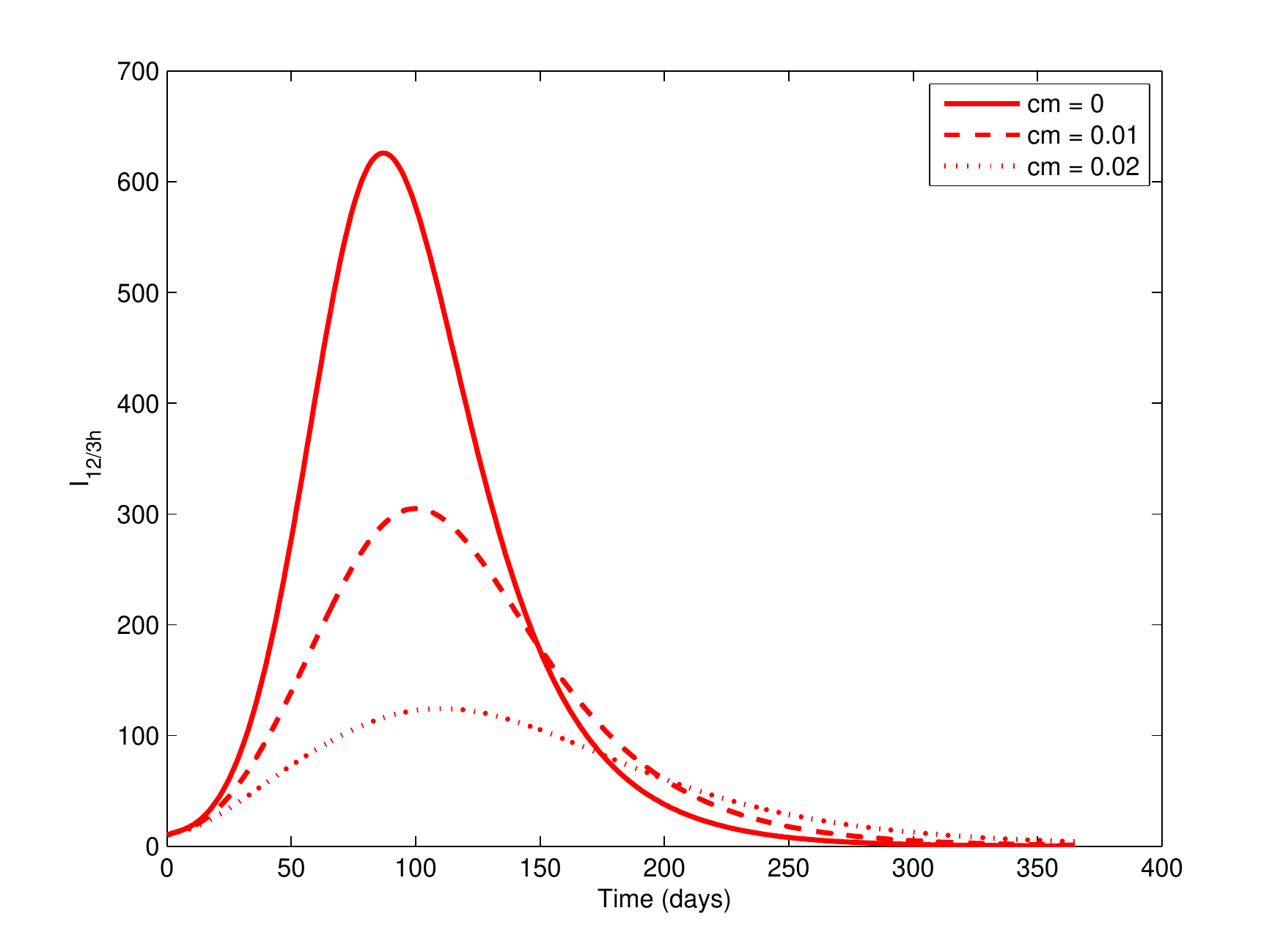}
\caption{$I_{1jh}$ (infected by DEN-$1$, then by DEN-$j$)}
\end{subfigure}
\begin{subfigure}[b]{0.45\textwidth}
\centering
\includegraphics[scale=0.41]{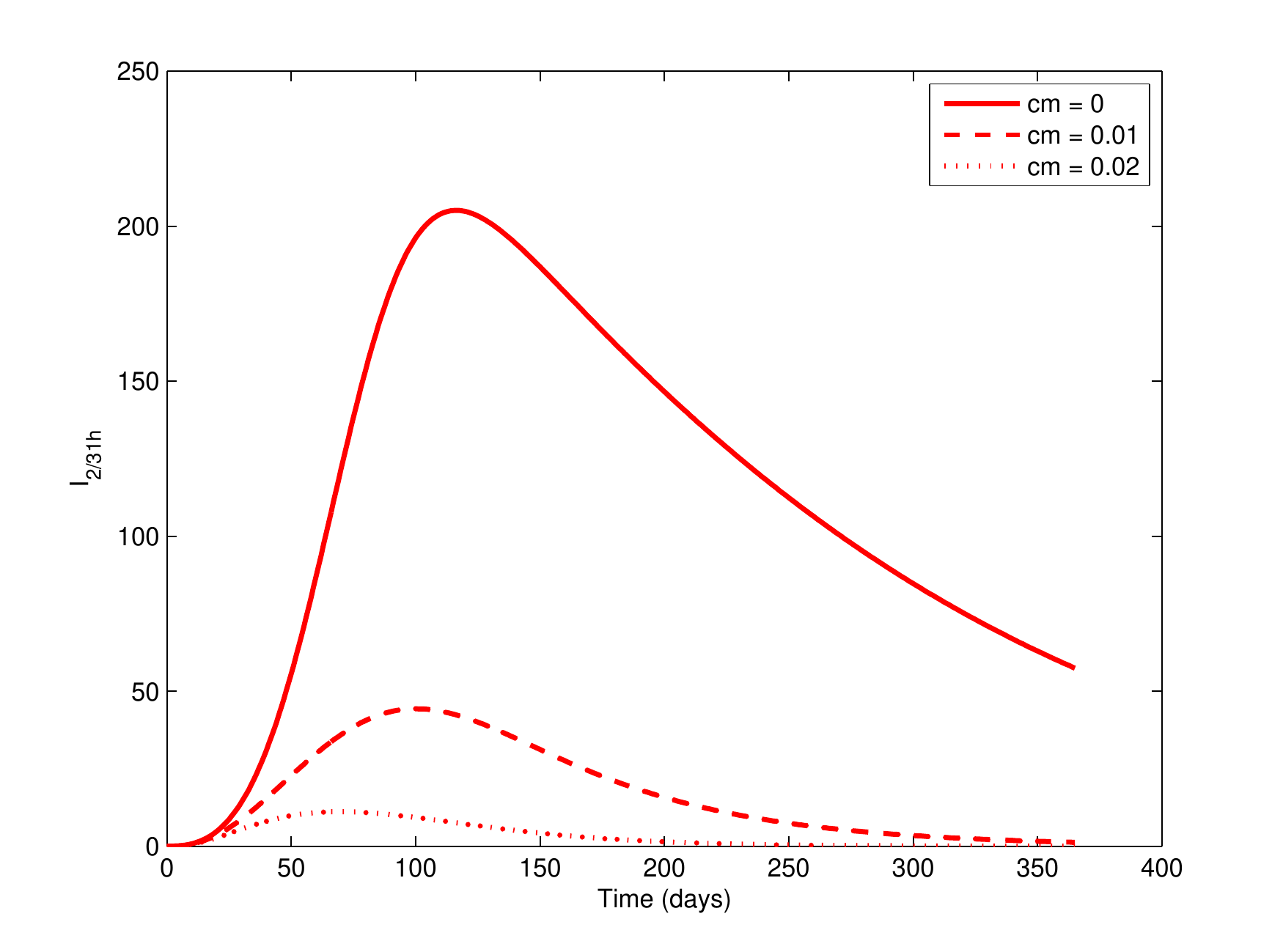}
\caption{$I_{j1h}$ (infected by DEN-$j$, then by DEN-$1$)}
\end{subfigure}
\caption{Infected individuals varying the control $c_m$ in case of two serotypes:
DEN-$1$ and DEN-$j$, $j\in \{2, 3\}$ (results for 1 year with
$\mu_{dhf}=0.02$,
$\sigma=1.1$,
$\beta_{1mh}=\beta_{1hm}=0.25$,
$\beta_{jmh}=\beta_{jhm}=0.33$,
$\eta_{1h}=1/7$,
$\eta_{jh}=1/9$,
and $I_{1h0} = I_{jh0} = I_{1jh0} = 10$)}
\label{var_cmbeta033sigma11}
\end{figure}

\begin{figure}
\centering
\begin{subfigure}[b]{0.45\textwidth}
\centering
\includegraphics[scale=0.41]{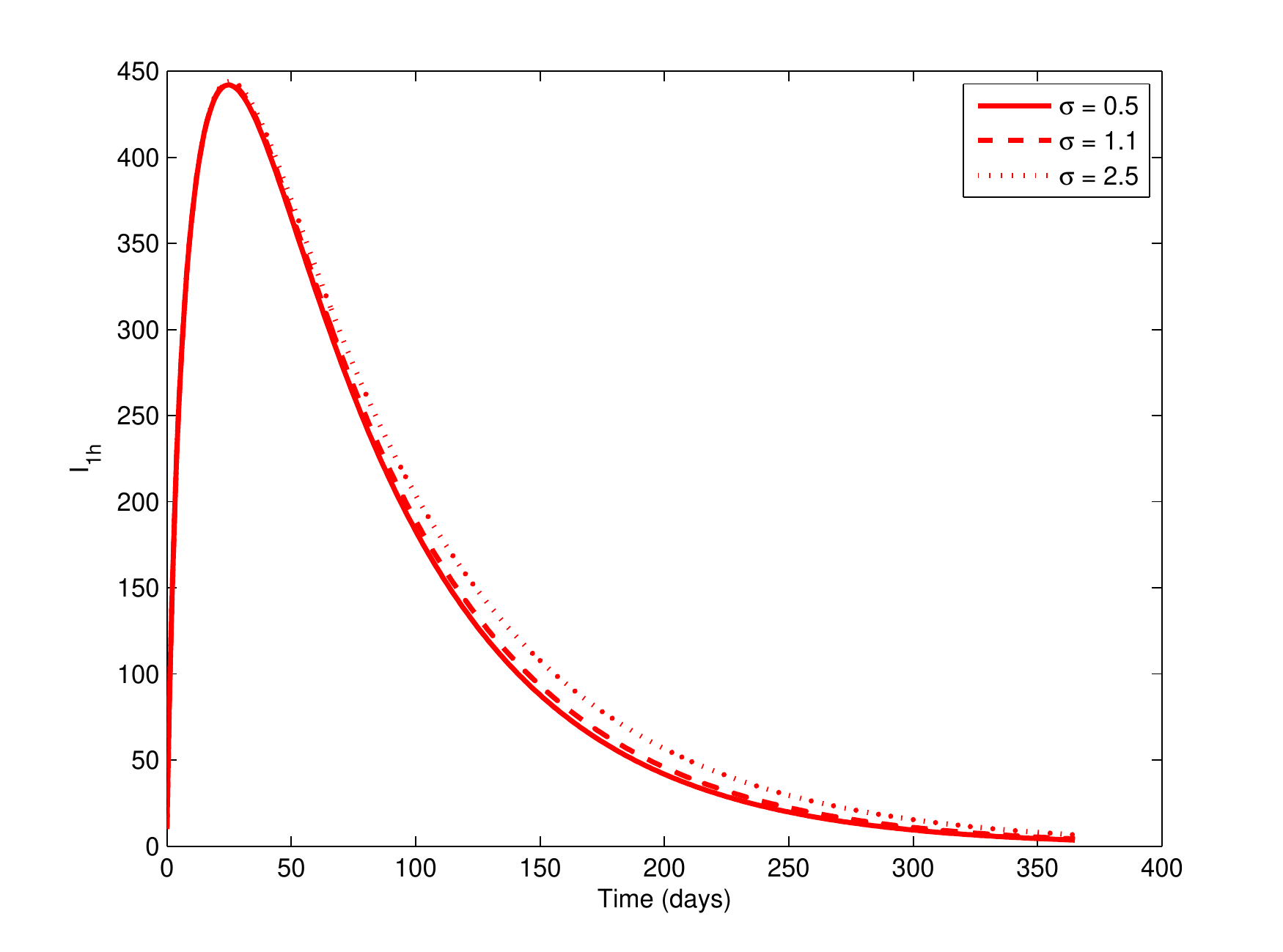}
\caption{$I_{1h}$}
\end{subfigure}%
\begin{subfigure}[b]{0.45\textwidth}
\centering
\includegraphics[scale=0.41]{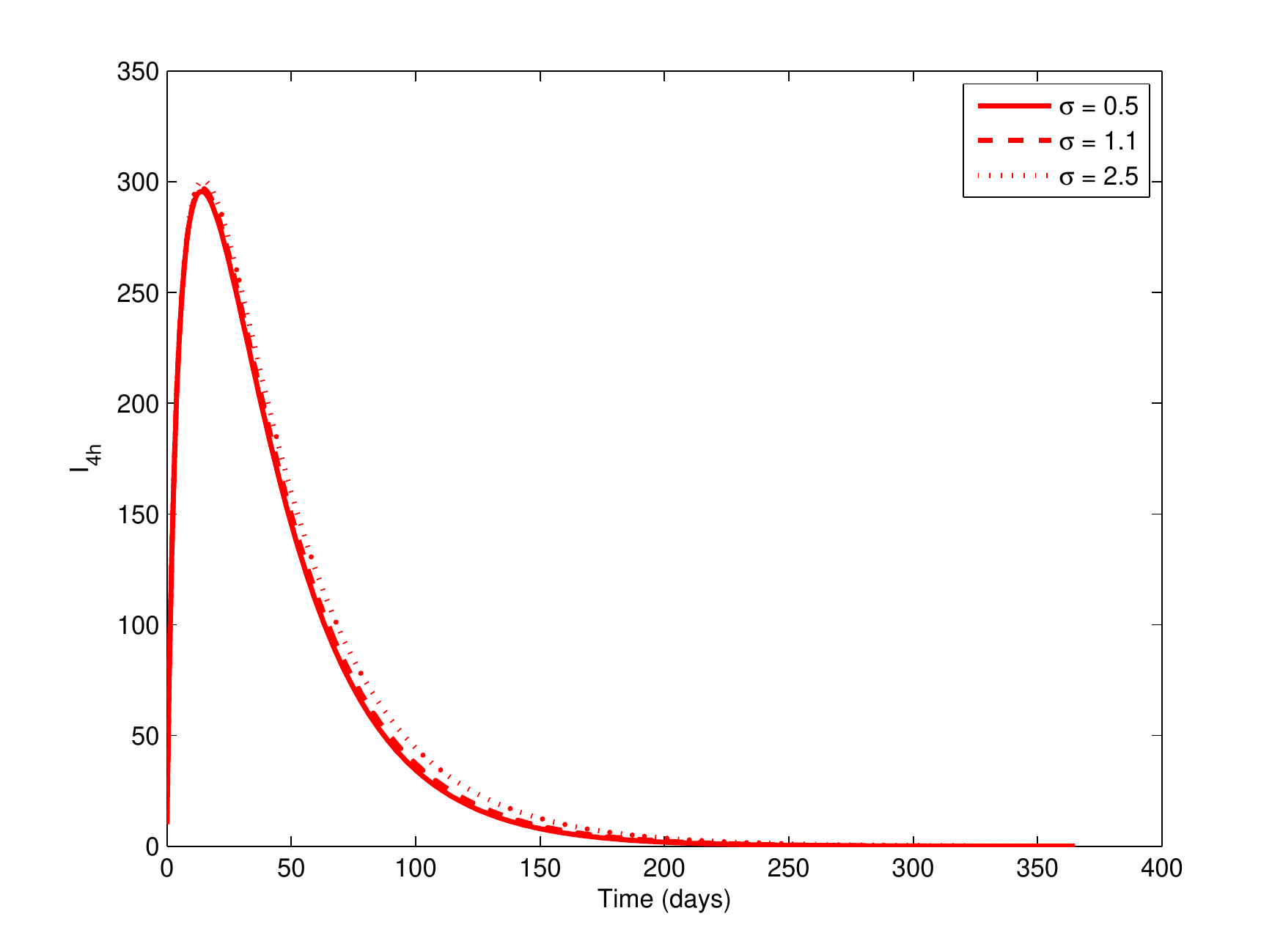}
\caption{$I_{jh}$}
\end{subfigure}\\
\begin{subfigure}[b]{0.45\textwidth}
\centering
\includegraphics[scale=0.41]{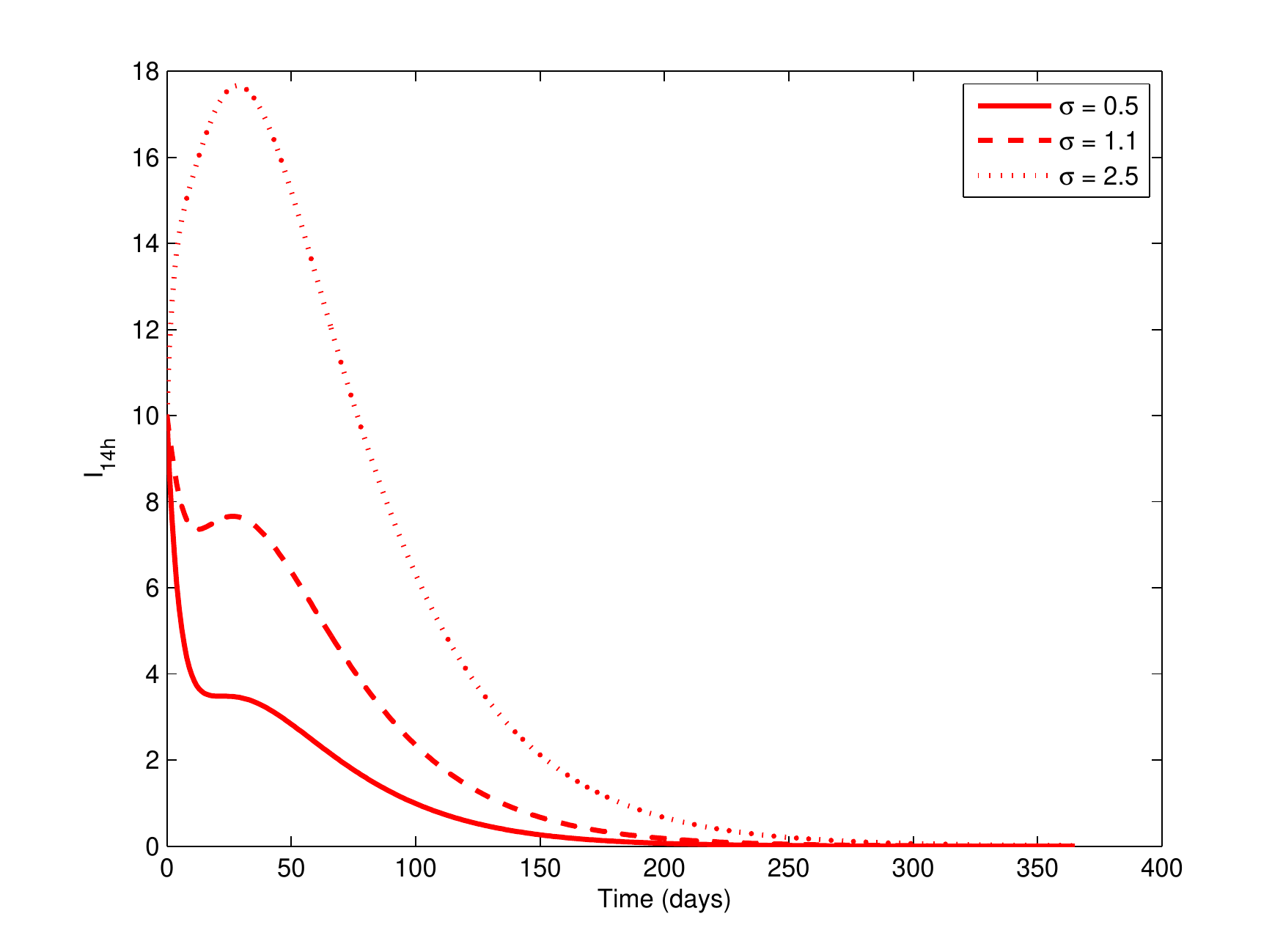}
\caption{$I_{1jh}$}
\end{subfigure}
\begin{subfigure}[b]{0.45\textwidth}
\centering
\includegraphics[scale=0.41]{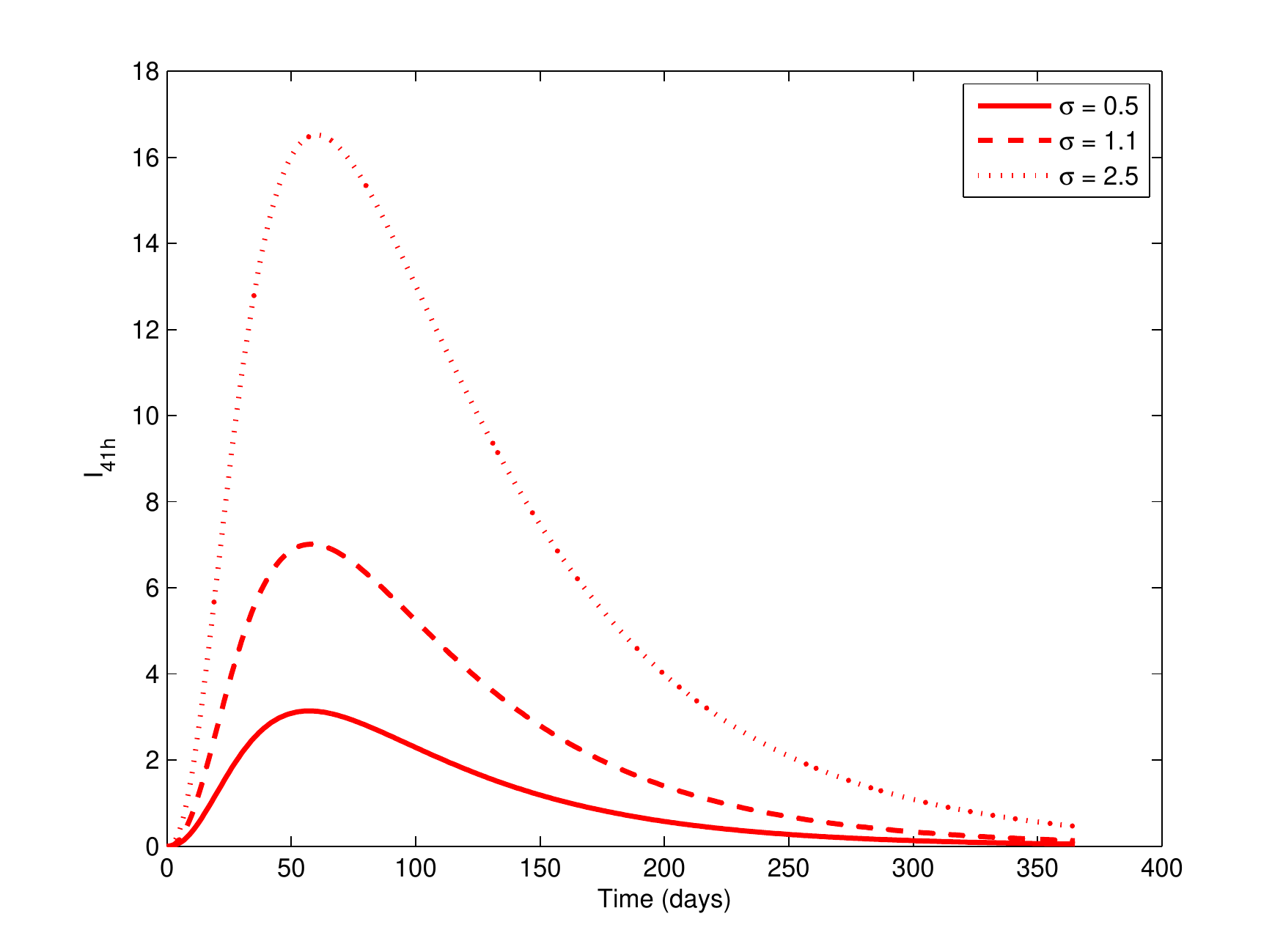}
\caption{$I_{j1h}$}
\end{subfigure}
\caption{Infected individuals varying $\sigma$,
in case of two serotypes DEN-$1$ and DEN-$4$}
\label{var_sigmabeta025cm001}
\end{figure}

One important information is the evolution of the human population $N_h$.
Figure~\ref{var_nh1} presents this evolution considering $\beta=0.25$ and
$\sigma=1.1$ for different controls. Using no insecticide, 450 deaths
are expected at the end of 154 days. However, the application of a small quantity
of insecticide reduces this number to 50.

We now increase $\beta$ to $0.33$ keeping $\sigma=1.1$ and varying
the control $c_m$ to study the influence of the transmission probability
(see Figure~\ref{var_cmbeta033sigma11}). As expected, the number of infected people rises,
having increased almost four-fold. It is thus concluded that $\beta$ is a very
sensitive parameter. As before, the impact of the control is crucial.

Figure~\ref{var_nh2} reports the evolution of the total population $N_h$ using
$\beta=0.33$ and $\sigma=1.1$ for different controls. In this figure the application
of insecticide is even more evident: with no control, human population can suffer
a significant decline. With $c_m = 0.02$, the fluctuation of total population
is almost non-existing. Comparing with Figure~\ref{var_nh1}, it is also noticeable
the influence of $\beta$. Increasing the transmission probability
from infected individuals, the population fall is bigger.
Note that $N_h$ starts decreasing at $t=40$,
about the same time the number of secondary infections become significant
(Figures~\ref{var_cmbeta033sigma11}(c) and (d)).

In the second phase of the numerical experiments, we fixed
$\beta=0.25$ and $c_m=0.01$ and tested the model varying $\sigma$
(see Figure~\ref{var_sigmabeta025cm001}). In Figures~\ref{var_sigmabeta025cm001}(a)
and \ref{var_sigmabeta025cm001}(b) the influence of the parameter $\sigma$
is residual. Although the ADE phenomenon is just related
to a secondary infection, $\sigma$ slightly affects the number of infected
mosquitos and, consequently, has an impact, even though small,
on the first infection. For this reason, this small influence only appears
a few days after the beginning of the outbreak. In Figures~\ref{var_sigmabeta025cm001}(c)
and \ref{var_sigmabeta025cm001}(d) the ADE phenomenon is considerable.
Increasing $\sigma$ from $0.5$ to $2.5$ leads to a five-fold increase
of second infections ($I_{1jh}$ and $I_{j1h}$).
\begin{figure}
\centering
\begin{subfigure}[b]{0.45\textwidth}
\centering
\includegraphics[scale=0.41]{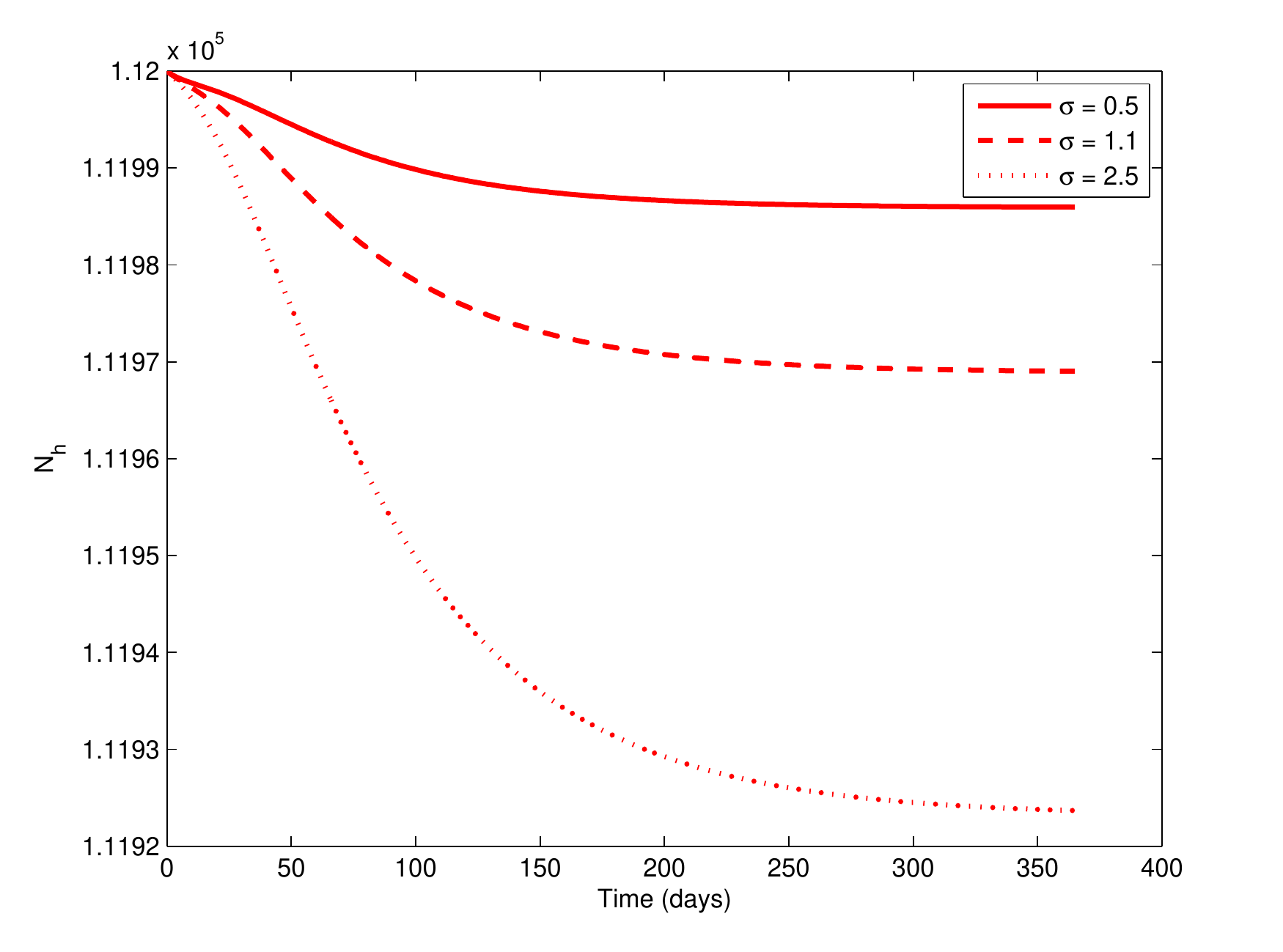}
\caption{DEN-$1$ and DEN-$4$}
\label{var_nh3}
\end{subfigure}%
\begin{subfigure}[b]{0.45\textwidth}
\centering
\includegraphics[scale=0.41]{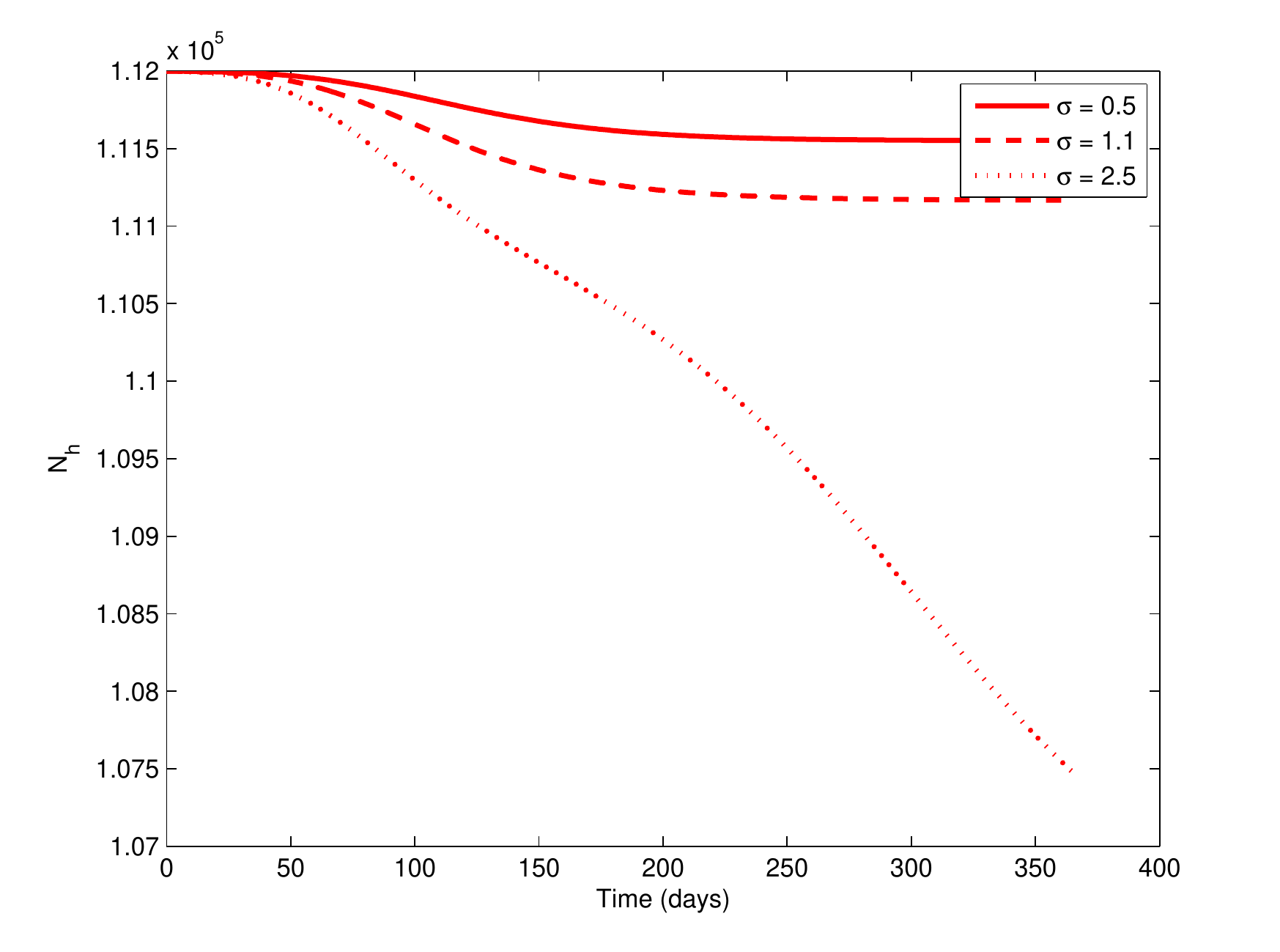}
\caption{DEN-$1$ and DEN-$j$, $j\in \{2, 3\}$}
\label{var_nh4}
\end{subfigure}
\caption{Total population $N_{h}$ varying $\sigma$,
in case of two serotypes}
\label{Fig810}
\end{figure}
\begin{figure}
\centering
\begin{subfigure}[b]{0.45\textwidth}
\centering
\includegraphics[scale=0.41]{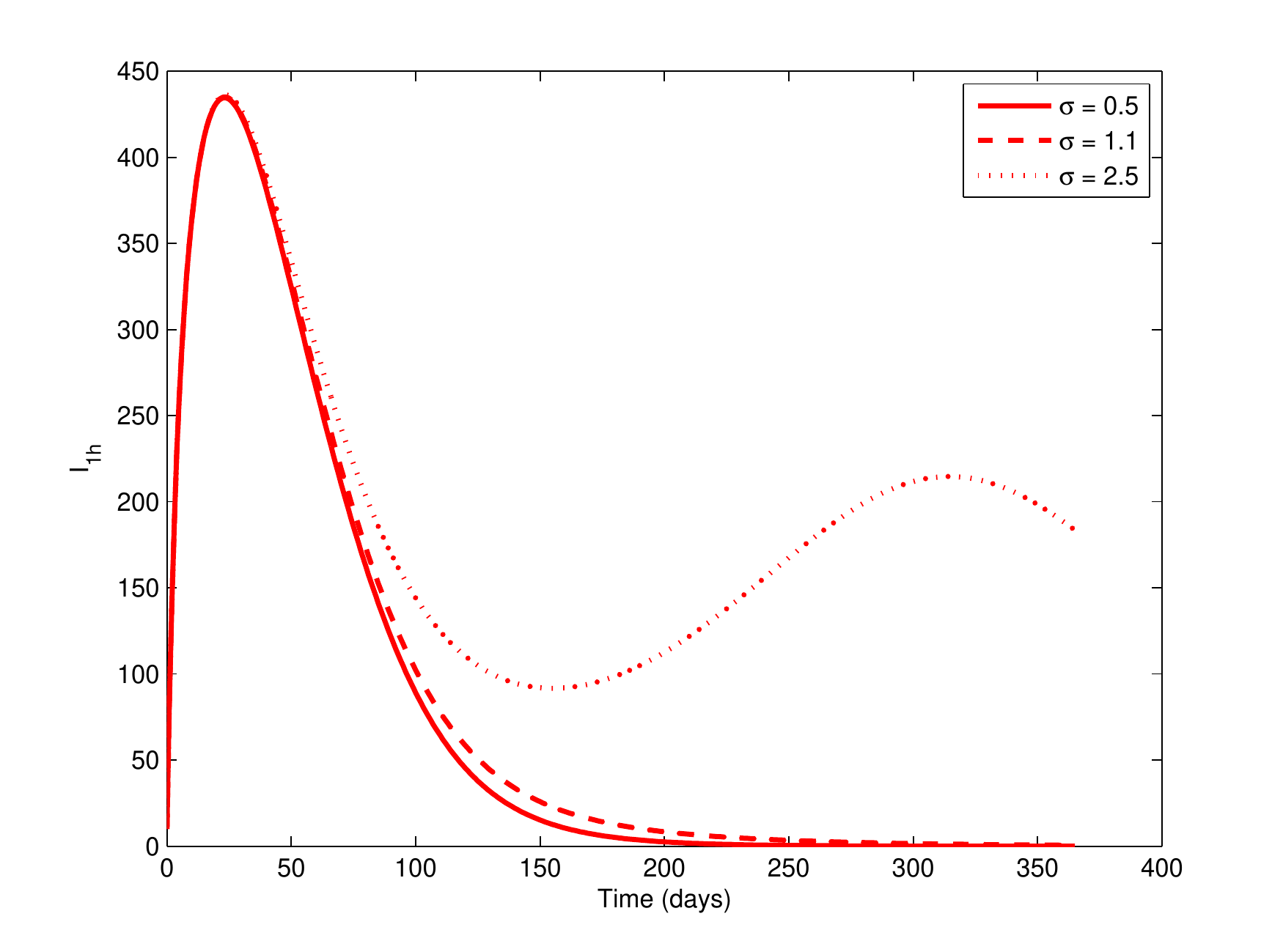}
\caption{$I_{1h}$}
\end{subfigure}%
\begin{subfigure}[b]{0.45\textwidth}
\centering
\includegraphics[scale=0.41]{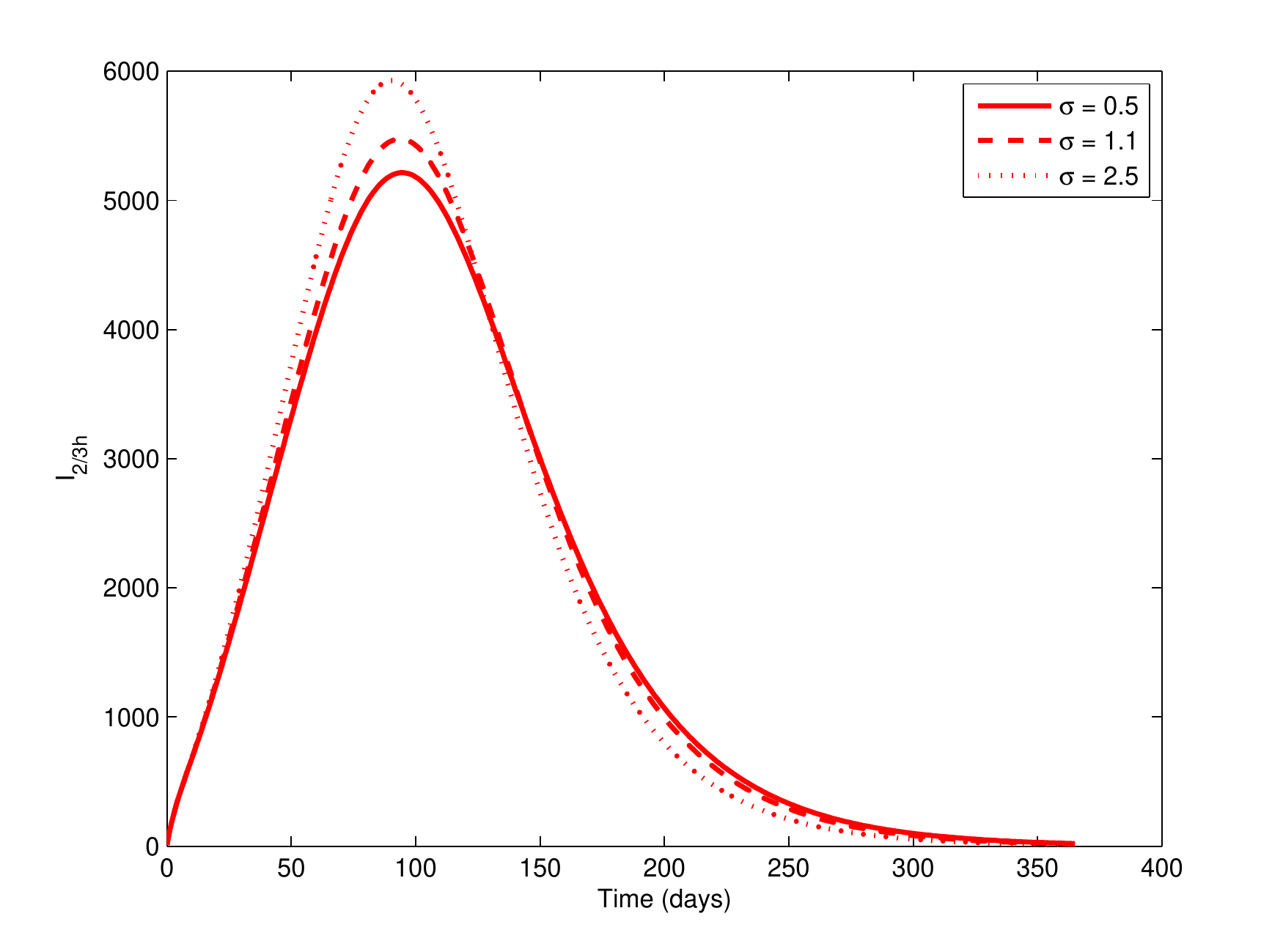}
\caption{$I_{jh}$}
\end{subfigure}\\
\begin{subfigure}[b]{0.45\textwidth}
\centering
\includegraphics[scale=0.41]{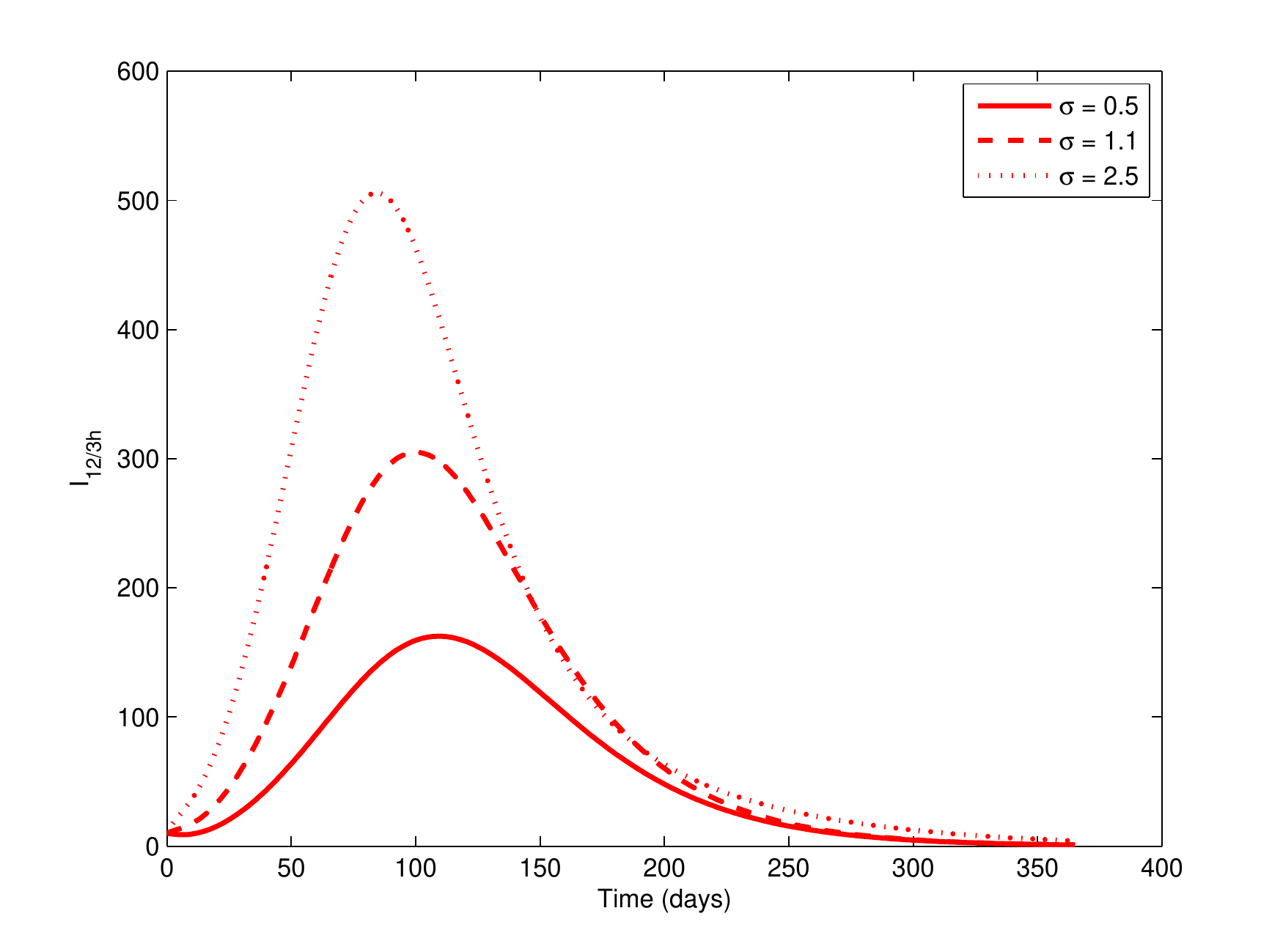}
\caption{$I_{1jh}$}
\end{subfigure}
\begin{subfigure}[b]{0.45\textwidth}
\centering
\includegraphics[scale=0.41]{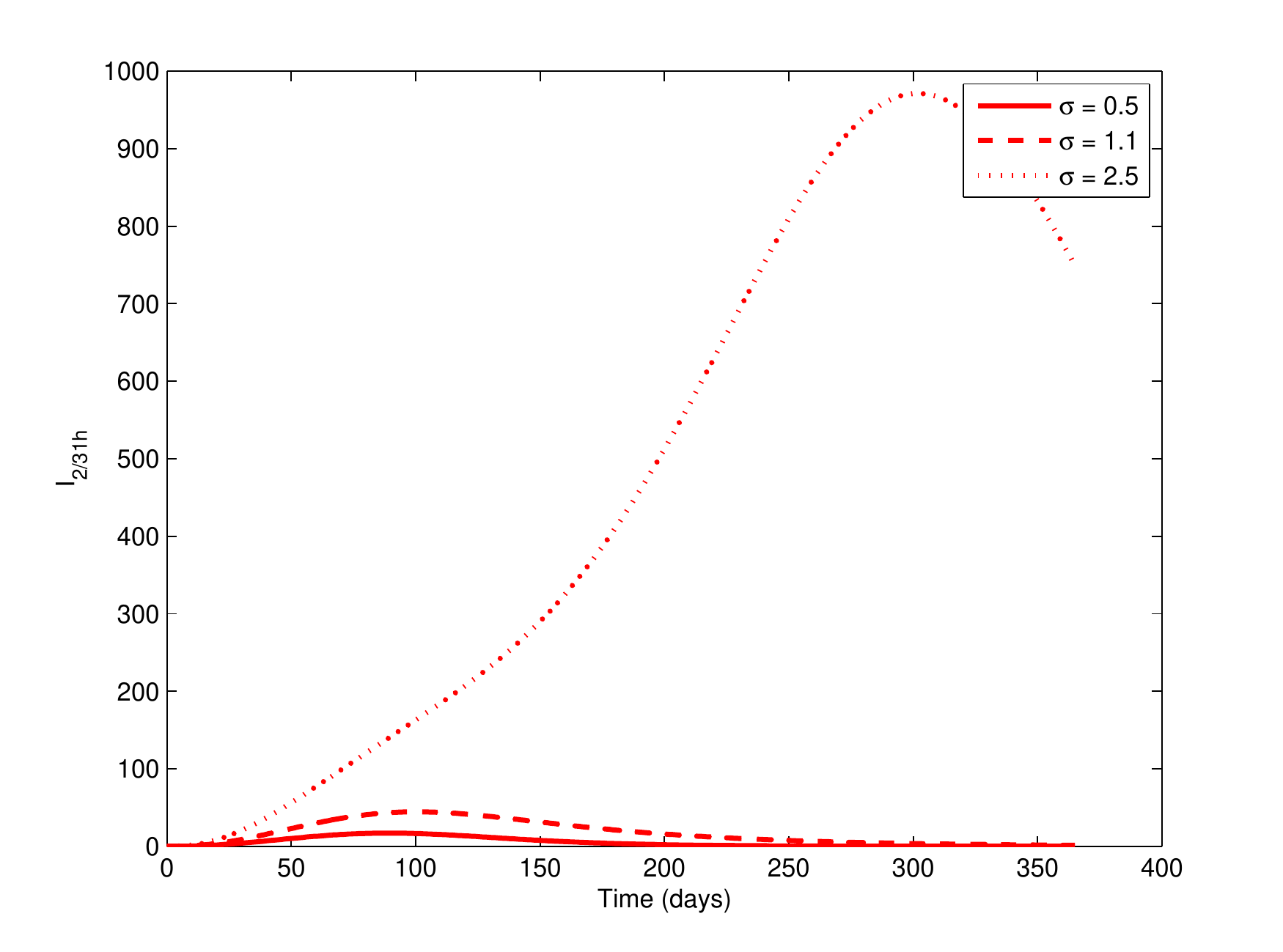}
\caption{$I_{j1h}$}
\end{subfigure}
\caption{Infected individuals varying $\sigma$,
in case of two serotypes DEN-$1$ and DEN-$j$, $j\in \{2, 3\}$}
\label{var_sigmabeta033cm001}
\end{figure}

Figures~\ref{Fig810} and \ref{var_sigmabeta033cm001}
show the results for $\beta$ increased to $0.33$ while keeping
the control $c_m=0.01$ and varying $\sigma$.
This allows to learn about the influence of the transmission probability $\beta$
($\sigma$ seems to be more sensitive with $\beta$ at $0.33$). On one hand,
it is evident in Figure~\ref{var_sigmabeta033cm001} the influence of ADE phenomenon.
On the other hand, when compared to Figure~\ref{var_sigmabeta025cm001},
it is obvious the importance of the $\beta$ parameters.

\begin{figure}
\centering
\begin{subfigure}[b]{0.45\textwidth}
\centering
\includegraphics[scale=0.49]{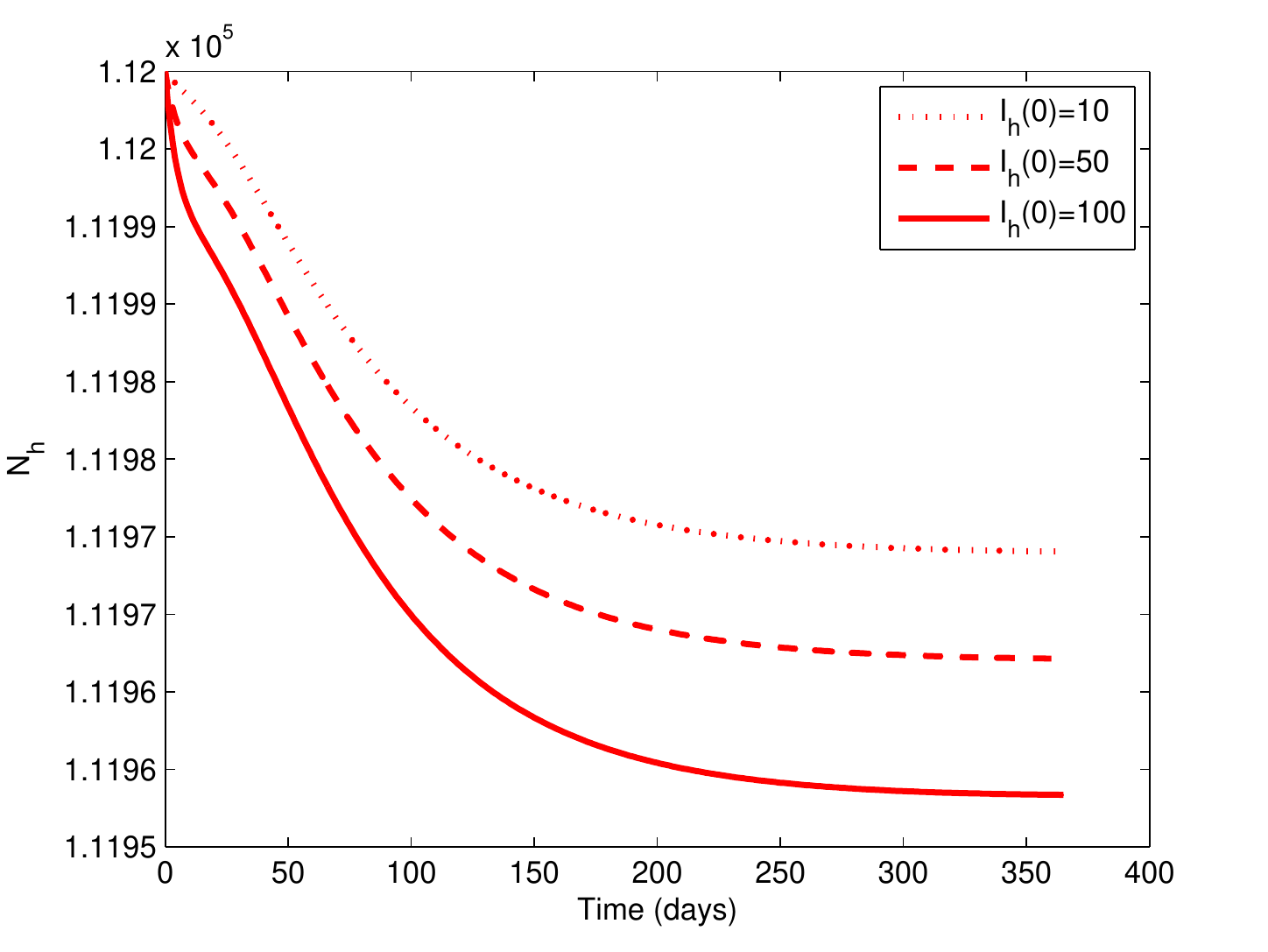}
\caption{DEN-$1$ and DEN-$4$\newline
($\beta_{jmh}=\beta_{jhm}=0.25$, $\eta_{jh}=1/5$)}
\end{subfigure}%
\begin{subfigure}[b]{0.45\textwidth}
\centering
\includegraphics[scale=0.41]{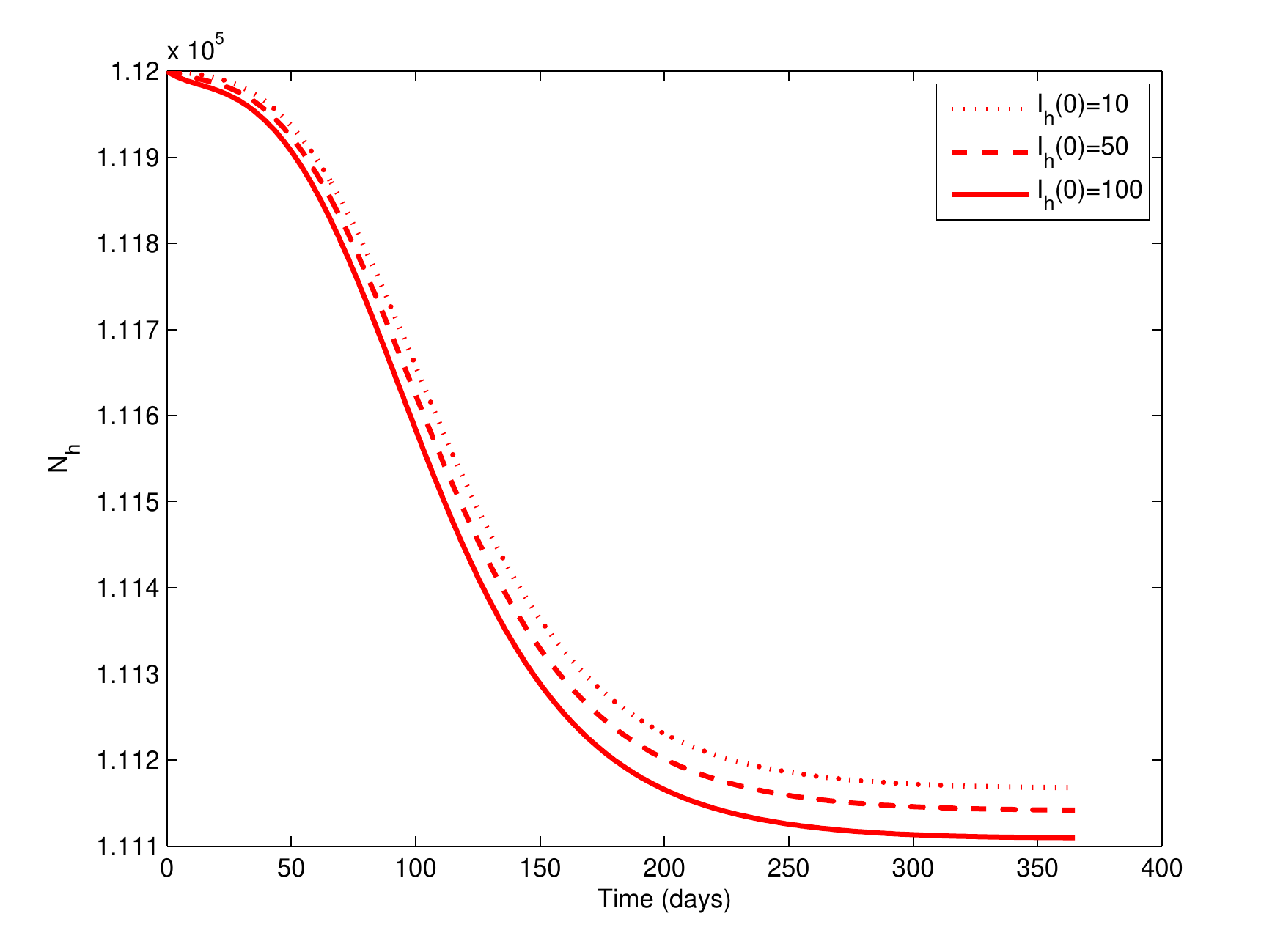}
\caption{DEN-$1$ and DEN-$j$, $j \in \{2, 3\}$\newline
($\beta_{jmh}=\beta_{jhm}=0.33$, $\eta_{jh}=1/9$)}
\end{subfigure}
\caption{Total population $N_h(t)$ varying the initial conditions
$I_{1h}(0) = I_{jh}(0) = I_{1jh}(0) = \xi$,
$\xi \in \{10, 50, 100\}$, in presence of
two serotypes (results for 1 year with
$\mu_{dhf}=0.02$,
$c_m = 0.01$,
$\sigma=1.1$,
$\beta_{1mh}=\beta_{1hm}=0.25$,
$\eta_{1h}=1/7$)}
\label{Nh_ci}
\end{figure}

\begin{figure}
\center
\includegraphics[scale=0.5]{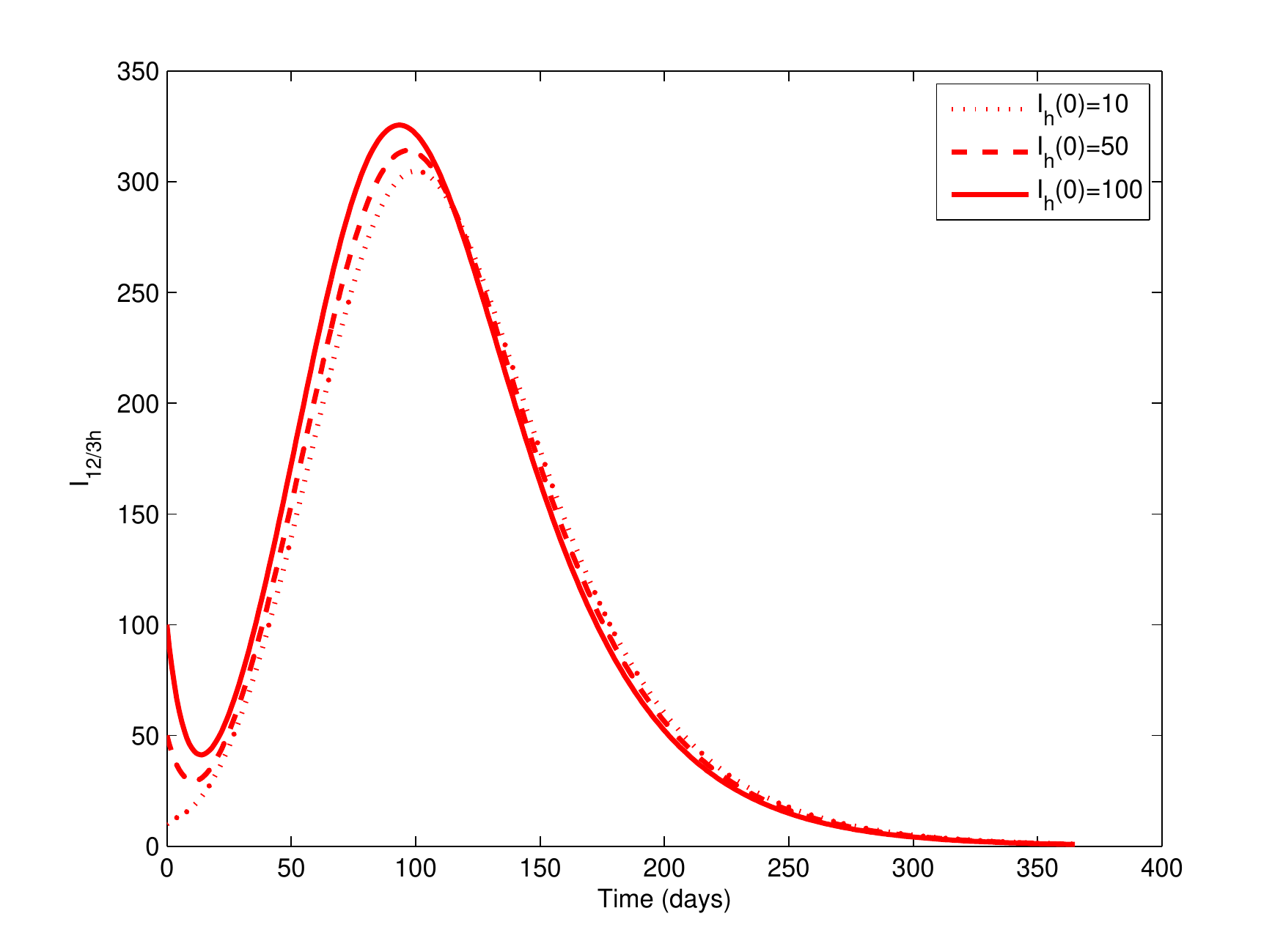}
\caption{Infected individuals first by DEN-1
and after by DEN-$j$, $j \in \{2, 3\}$
(results for 1 year with
$\mu_{dhf}=0.02$,
$c_m = 0.01$,
$\sigma=1.1$,
$\beta_{1mh}=\beta_{1hm}=0.25$,
$\beta_{jmh}=\beta_{jhm}=0.33$,
$\eta_{1h}=1/7$,
and $\eta_{jh}=1/9$)}
\label{I1j_ci_j23}
\end{figure}

Additionally, we also study how many initial
infections of new serotypes would allow these
serotypes to persist in the population if no control measures were to be taken.
For this, we set the value $c_m = 0$ and considered
different initial values of infected people by the new serotype.
More precisely, at time zero we considered $I_{1h}(0) = I_{jh}(0) = I_{1jh}(0) = \xi$
with $\xi \in \{10, 50, 100\}$. For both serotypes DEN-2/3 or DEN-4
the curves have almost the same shape, the small variations in value being
the expected ones: the curves $I_{1h}(t)$, $I_{14h}(t)$, $I_{4h}(t)$, $I_{41h}(t)$,
$R_{14h}(t)$, $R_{4h}(t)$, $R_{41h}(t)$, $I_{2/3h}(t)$, $I_{2/31h}(t)$, $R_{1h}(t)$,
$R_{12/3h}(t)$, $R_{2/3h}(t)$, $R_{2/31h}(t)$
increase continuously with the increase of the initial value $\xi$;
the curve $N_h(t)$ decrease continuously with the increase of the initial value $\xi$:
see Figure~\ref{Nh_ci}.
Curiously, as can be seen in Figure~\ref{I1j_ci_j23},
an exception in behavior occurs in the first month
for the curve $I_{1jh}(t)$, $j \in \{2, 3\}$.

\begin{figure}
\center
\includegraphics[scale=0.5]{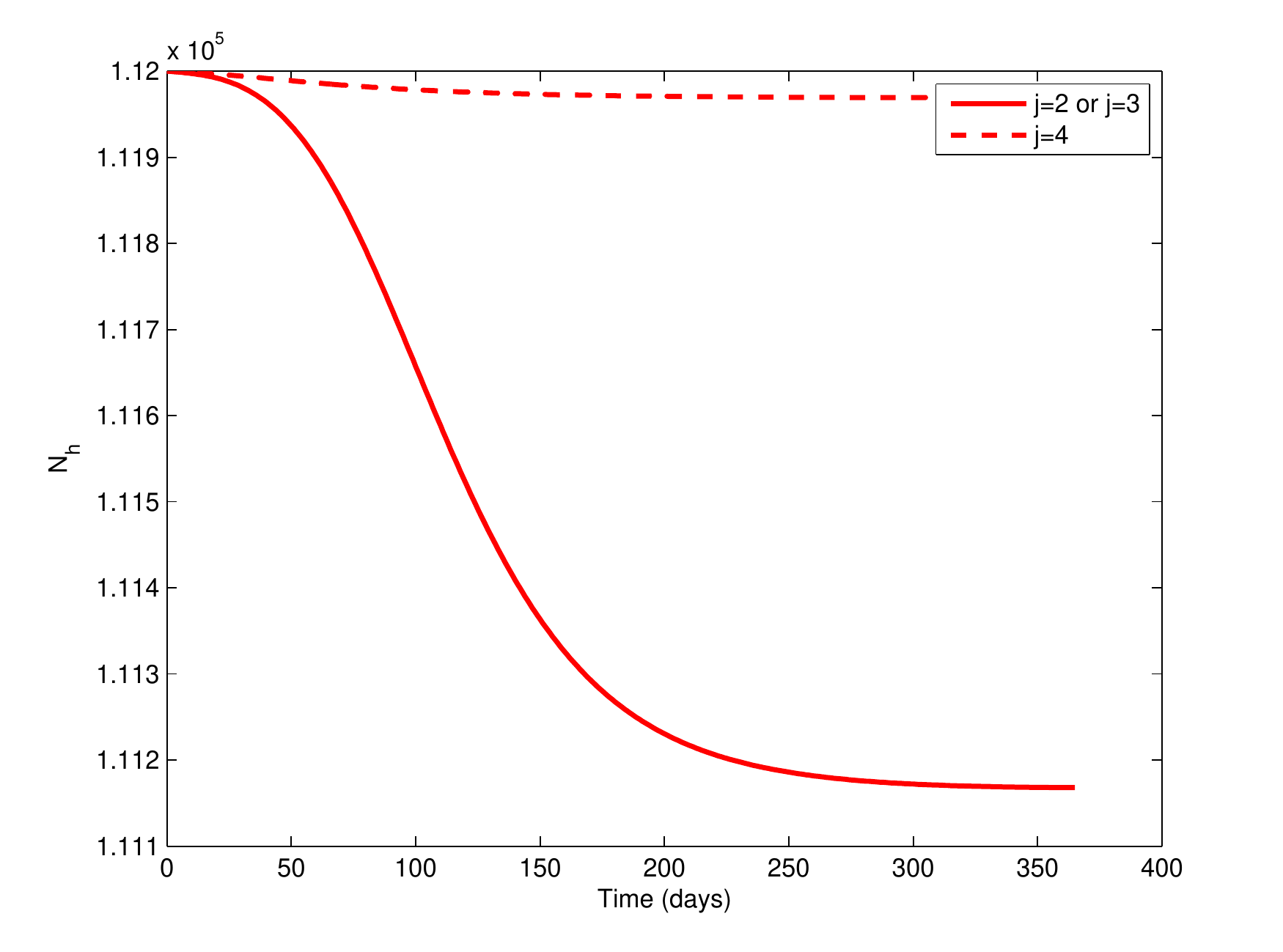}
\caption{Total population $N_h$
($\mu_{dhf}=0.02$,
$c_m = 0.01$,
$\sigma=1.1$,
$\beta_{1mh}=\beta_{1hm}=0.25$,
$\eta_{1h}=1/7$,
and $I_{1h0} = I_{jh0} = I_{1jh0} = 10$)
for
DEN-$4$ ($\beta_{jmh}=\beta_{jhm}=0.25$, $\eta_{jh}=1/5$)
versus DEN-$j$, $j\in \{2, 3\}$
($\beta_{jmh}=\beta_{jhm}=0.33$, $\eta_{jh}=1/9$)}
\label{D23moreAgD4}
\end{figure}

We finalize by remarking that the scales of figures for scenario
$j \in \{2, 3\}$ (serotype more aggressive than DEN-1)
and for scenario $j = 4$ (serotype less aggressive than DEN-1) are quite
different. Figure~\ref{D23moreAgD4} illustrates well how much more aggressive
DEN-2 or DEN-3 is than DEN-4.


\section{Conclusion}
\label{sec:5}

According to the World Health Organization (WHO), at least
40\% of the world population is at risk from dengue and several outbreaks have
appeared recently in Europe. The first was recorded in Greece and consisted
of two serotypes, Madeira's outbreak was the second epidemics of dengue in Europe.
Local authorities in Madeira continue to monitor Aedes aegypti entomological
activity through a program of traps \cite{DGS,carla2012}.
Moreover, the report \cite{ECDC}
of the European Centre for Disease Prevention and Control (ECDC)
concludes that ``The important increase of co-circulation
of different dengue virus serotypes worldwide is a potential
source of its re-introduction to Madeira in the future'' and
``The possibility of sustainability of the virus during
the winter season and subsequent reemergence when climatic conditions
again favor Aedes aegypti mosquito activity exists''.
Motivated by these, we proposed an epidemic model for dengue fever infection
that considers the simultaneous spreading of two different
serotypes throughout shared non-naive host population,
with different virulence and transmission intensity.
Its dynamics was discussed in the context of Madeira island,
and the influence of the infected transmission probability
(human and mosquito) analyzed.
The Antibody-Dependent Enhancement (ADE), which results from
a new infection with a particular serotype in an
individual with acquired immunity to a different serotype, was also studied.
The higher the value of ADE, the higher is the number of infected individuals.
More important, while classical dengue fever causes negligible mortality,
ADE leads to the development of a significantly more dangerous dengue.
The important role of insecticide as a control measure,
is also a relevant conclusion.

This study is useful for strategies of health control on Madeira Island.
For example, we used one control variable related with application of insecticide.
It turns out that application of insecticide was done in Madeira intensively,
mainly carbamate formulations (bendiocarb) and
Bacillus thuringiensis israelensis (bti), which is a bacteria-derived toxin
that is lethal when ingested by larvae \cite{Bhatt2013}.
Indoor spraying at the biting time of the Aedes aegypti mosquito (early morning
or late afternoon) is also easily done by population
(dwelling rooms were sprayed by inhabitants of Madeira Island at the
end of the day with doors closed, e.g., bathrooms, kitchens and toilets).
However, the model investigated here is sufficiently general
to be applied in other contexts.


\section*{Acknowledgements}

This work was supported by \emph{The Portuguese Foundation
for Science and Technology} (FCT): Rodrigues and Torres through
CIDMA within project UID/MAT/04106/2013;
Monteiro by the ALGORITMI R\&D Center and project
UID/CEC/00319/2013; Torres by project PTDC/EEI-AUT/1450/2012,
co-financed by FEDER under POFC-QREN with COMPETE reference
FCOMP-01-0124-FEDER-028894. The authors are grateful to
Alexandre Arag\~{a}o for improving the English;
and to two referees for several constructive comments and remarks.


\small




\begin{thebibliography}{00}

\bibitem{Adams2006}
B.~Adams, E.~C. Holmes, C.~ Zhang, M.~P.~Mammen, Jr., S.~Nimmannitya, S.~Kalayanarooj and M.~Boots.
\newblock {\em Cross-protective immunity can account for the alternating
epidemic pattern of dengue virus serotypes circulating in Bangkok}.
\newblock {\em PNAS} 103(38):14234--14239, 2006.

\bibitem{Anderson}
R.~M. Anderson and R.~M. May.
\newblock {\em Infectious diseases of humans: Dynamics and control}.
\newblock Oxford University Press, 1991.

\bibitem{Balmaseda}
A. Balmaseda \emph{et al.}
\newblock Serotype-specific differences in clinical manifestations of dengue.
\newblock {\em Am. J. Trop. Med. Hyg.}, 74(3):449--456, 2006.

\bibitem{Barrios2013}
J.~Barrios, A.~Pi\'{e}trius, G.~Joya, A.~Marrero and H.~de Arazoza.
\newblock A differential inclusion approach for modeling and analysis
of dynamical systems under uncertainty. Application to dengue disease transmission.
\newblock {\em Soft Comput}, 17:239--253, 2013.

\bibitem{Bhatt2013}
S. Bhatt \emph{et al.}
\newblock The global distribution and burden of dengue.
\newblock {\em Nature}, 496:504--507, 2013.

\bibitem{Bianco2009}
S.~Bianco, L.~B.~Shaw and I.~B.~Schwartz
\newblock Epidemics with multistrain interactions: the interplay
between cross immunity and antibody-dependent enhancement.
\newblock {\em Chaos}, 19(4):043123, 2009.

\bibitem{Cattand2006}
P.~Cattand \emph{et al.}
\newblock Tropical Diseases Lacking Adequate Control Measures:
Dengue, Leishmaniasis, and African Trypanosomiasis.
\newblock {\em Disease Control Priorities in Developing Countries},
\newblock DCPP Publications, 451--466, 2006.

\bibitem{CDC}
{C}{D}{C}. Centers for Diseases Control and Prevention
\newblock Dengue Fever (DF).
\newblock \url{http://www.cdc.gov/dengue/faqFacts/fact.html}.

\bibitem{Chan2012}
M.~Chan and M.~A.~ Johansson.
\newblock The Incubation Periods of Dengue Viruses.
\newblock {\em PLoS ONE}, 7(11):e50972, 2012.

\bibitem{DengueNet}
{Dengue {V}irus {N}et}.
\newblock \url{http://denguevirusnet.com}, 2013.

\bibitem{Diekmann}
O. Diekmann and J.~A.~P. Heesterbeek.
\newblock {\em Mathematical epidemiology of infectious diseases:
Model building, analysis and interpretation}.
\newblock New York, John Wiley and Sons, 2000.

\bibitem{DGS}
{D}{G}{S}. Directorate-General of Health.
\newblock Dengue Madeira.
\newblock \url{http://www.dgs.pt}.

\bibitem{ECDC}
ECDC.
\newblock Dengue outbreak in Madeira, Portugal, October–November 2012.
\newblock European Centre for Disease Prevention and Control, Stockholm, 2013.

\bibitem{Esteva2003}
L.~Esteva and C.~Vargas
\newblock Coexistence of different serotypes of dengue virus,
\newblock {\em J. Math. Biol.}, 46:31--47, 2003.

\bibitem{Feng1997}
Z.~Feng, J.~X.~Velasco-Hernández
\newblock Competitive exclusion in a vector-host model for the dengue fever,
\newblock {\em J. Math. Biol.}, 35:523--544, 1997.

\bibitem{Focks2000}
D.~A. Focks, R.~J.~Brenner, J.~Hayes and E.~Daniels.
\newblock Transmission thresholds for dengue in terms of \emph{Aedes aegypti}
pupae per person with discussion of their utility in source reduction efforts.
\newblock {\em Am. J. Trop. Med. Hyg.}, 62:11--18, 2000.

\bibitem{Focks1995}
D.~A. Focks, E.~Daniels, D.~G.~Haile, J.~E.~Keesling.
\newblock A simulation model of the epidemiology of urban dengue fever:
literature analysis, model development, preliminary validation, and
samples of simulation results.
\newblock {\em Am. J. Trop. Med. Hyg.}, 53:489--506, 1995.

\bibitem{Focks1993}
D.~A.~Focks, D.~G.~Haile, E.~Daniels and G.~A.~Mount
\newblock Dynamic life table model for \emph{Aedes aegypti} (Diptera: Culicidae):
analysis of the literature and model development
\newblock {\em J. Med. Entomol.}, 30:1003--1017, 1993.

\bibitem{Gjenero-Margan2011}
I.~Gjenero-Margan \emph{et al.}
\newblock Autochthonous dengue fever in {C}roatia, {A}ugust-{S}eptember 2010.
\newblock {\em Euro Surveill.} 16(9), 2011.

\bibitem{Harrington2001}
L.~C.~Harrington \emph{et al.}
\newblock Analysis of survival of young and old \emph{Aedes aegypti}
(Diptera: Culicidae) from Puerto Rico and Thailand
\newblock {\em Journal of Medical Entomology}, 38:537--547, 2001.

\bibitem{censos2011}
{I}{N}{E}. Statistics Portugal.
\newblock  \url{http://censos.ine.pt}

\bibitem{INSA}
{I}{N}{S}{A}. National Health Institute Doutor Ricardo Jorge.
\newblock Dengue Madeira.
\newblock \url{http://www.insa.pt/sites/INSA/Portugues/ComInf/Noticias/Paginas/DengueMadeiraDiagLab.aspx},
2012.

\bibitem{Ruche2010}
G. La Ruche \emph{et al.}
\newblock First two autochthonous dengue virus infections in metropolitan {F}rance, {S}eptember 2010.
\newblock {\em Euro Surveill.} 15(39), 2010.

\bibitem{Luz2003}
P.~M.~Luz, C.~T. Code\c{c}o, E.~Massad and C.~J.~Struchiner.
\newblock Uncertainties Regarding Dengue Modeling in {R}io de {J}aneiro, {B}razil.
\newblock {\em Mem. Inst. Oswaldo Cruz}, 98(7):871--878, 2003.

\bibitem{Freitas2007}
R.~Maciel-de-Freitas, W.~A.~Marques, R.~C.~Peres, S.~P.~Cunha
and R. Louren\c{c}o-de-Oliveira.
\newblock Variation in \emph{Aedes aegypti} (Diptera: Culicidae)
container productivity in a slum and a suburban district of
Rio de Janeiro during dry and wet seasons.
\newblock {\em Mem. Inst. Oswaldo Cruz}, 102:489--496, 2007.

\bibitem{Nisalak}
A.~Nisalak \emph{et al.}
\newblock Serotype-specific dengue virus circulation and dengue disease in Bangkok,
Thailand from 1973--1979.
\newblock {\em Am. J. Trop. Med. Hyg.}, 68: 191--202, 2003.

\bibitem{Nuraini2007}
N. Nuraini, E. Soewono and K.A. Sidarto,
\newblock Mathematical Model of Dengue Disease Transmission with Severe DHF Compartment,
\newblock {\em Bull. Malays. Math Sci. Soc}, 39(2):143--157, 2007.

\bibitem{Otero2008}
M.~Otero, N.~Schweigmann and H. G. Solari.
\newblock A stochastic spatial dynamical model for \emph{aedes aegypti}.
\newblock {\em Bull. Math. Biol.}, 70(5):1297--1325, 2008.

\bibitem{Padmanabla2011}
H.~Padmanabha, C.~C.~Lord and L.~P.~Lounibos.
\newblock Temperature induces trade-offs between development
and starvation resistance in \emph{Aedes aegypti} (L.) larvae.
\newblock {\em Med. Vet. Entomol.}, 25(4):445--453, 2011.

\bibitem{Sofia2009}
H.~S. Rodrigues, M.~T.~T. Monteiro and D.~F.~M. Torres.
\newblock Optimization of dengue epidemics:
a test case with different discretization schemes.
\newblock {\em AIP Conf. Proc.}, 1168(1):1385--1388, 2009.
{\tt arXiv:1001.3303}

\bibitem{Sofia2010c}
H.~S. Rodrigues, M.~T.~T. Monteiro and D.~F.~M. Torres.
\newblock Insecticide control in a dengue epidemics model.
\newblock {\em AIP Conf. Proc.}, 1281(1):979--982, 2010.
{\tt arXiv:1007.5159}

\bibitem{MyID:168}
H.~S. Rodrigues, M.~T.~T. Monteiro and D.~F.~M. Torres.
\newblock Dynamics of dengue epidemics when using optimal control.
\newblock {\em Math. Comput. Modelling}, 52(9-10):1667--1673, 2010.
{\tt arXiv:1006.4392}

\bibitem{Sofia2013}
H.~S. Rodrigues, M.~T.~T. Monteiro and D.~F.~M. Torres.
\newblock Dengue in {C}ape {V}erde: vector control and vaccination.
\newblock {\em Math. Popul. Stud.}, 20(4):208--223, 2013.
{\tt arXiv:1204.0544}

\bibitem{MyID:263}
H.~S. Rodrigues, M.~T.~T. Monteiro and D.~F.~M. Torres.
\newblock Bioeconomic perspectives to an optimal control dengue model.
\newblock {\em Int. J. Comput. Math.}, 90(10):2126--2136, 2013.
{\tt arXiv:1303.6904}

\bibitem{MyID:283}
H.~S. Rodrigues, M.~T.~T. Monteiro and D.~F.~M. Torres.
\newblock Vaccination models and optimal control strategies to dengue.
\newblock {\em Math. Biosci.}, 247: 1--12, 2014.
{\tt arXiv:1310.4387}

\bibitem{Sofia2014}
H.~S. Rodrigues, M.~T.~T. Monteiro, D.~F.~M. Torres, A.~C. Silva, C. Sousa and C. Concei\c{c}\~{a}o.
\newblock Dengue in Madeira Island.
\newblock In: {\em Mathematics of Planet Earth: Dynamics, Games and Science}
(Eds. J. P. Bourguignon, R. Jeltsch, A. Pinto and M. Viana),
CIM Series in Mathematical Sciences, Springer, Chapter~32,
DOI:10.1007/978-3-319-16118-1\_32
{\tt arXiv:1409.7915}

\bibitem{Sofia2012}
H.~S. Rodrigues, M.~T.~T. Monteiro, D.~F.~M. Torres and A. Zinober.
\newblock Dengue disease, basic reproduction number and control.
\newblock {\em Int. J. Comput. Math.}, 89(3):334--346, 2012.
{\tt arXiv:1103.1923}

\bibitem{tese_goncalo}
G.~F.~R. Seixas.
\newblock \emph{Aedes} (Stegomyia) \emph{aegypti} (Diptera, Culicidae) da ilha da Madeira:
origem geogr\'{a}fica e resist\^{e}ncia aos insecticidas.
\newblock {\em Master Thesis}, Universidade Nova de Lisboa, 2012.

\bibitem{Shuman}
E.~K. Shuman.
\newblock Global climate change and infectious diseases. Emerging Infectious Diseases.
\newblock {\em New England Journal of Medicine} 362:1061--1063, 2010.

\bibitem{carla2012}
C.~A.~Sousa \emph{et al.}
\newblock Ongoing outbreak of dengue type 1 in the
{A}utonomous {R}egion of {M}adeira, {P}ortugal: preliminary report.
\newblock {\em Euro Surveill} 17(49), 2012.

\bibitem{Stephenson2005}
J.~R.~Stephenson,
\newblock Understanding dengue pathogenesis: implications for vaccine design.
\newblock {\em Bull World Health Organ.} 83(4):308--314, 2005.

\bibitem{Driessche2002}
P.~van den Driessche and J.~Watmough.
\newblock Reproduction numbers and sub-threshold endemic
equilibria for compartmental models of disease transmission.
\newblock {\em Math. Biosci.}, 180: 29--48, 2002.

\bibitem{Vaughn}
D. W. Vaughn \emph{et al.}
\newblock Dengue viremia titer, antibody response pattern, and virus serotype
correlate with disease severity.
\newblock {\em J. Infect. Dis.} 181: 2--9, 2000.

\bibitem{Wahala2011}
W.~M.~P.~B. Wahala and A.~M. de Silva
\newblock The Human Antibody Response to Dengue Virus Infection.
\newblock {\em Viruses}, 3(12): 2374--2395, 2011.

\bibitem{Watson1999}
T.~M.~Watson and B.~H. Kay
\newblock Vector competence of \emph{Aedes} notoscriptus (Diptera: Culicidae)
for Barmah Forest Virus and of this species and \emph{Aedes aegypti}
(Diptera: Culicidae) for dengue 1-4 viruses in Queensland, Australia.
\newblock {\em Journal of Medical Entomology}, 36:508--514, 1999.

\bibitem{Wearing2006}
H.~J. Wearing.
\newblock Ecological and immunological determinants of dengue epidemics.
\newblock {\em Proc. Natl. Acad. Sci {USA}}, 103(31):11802--11807, 2006.

\bibitem{Who2009}
{WHO}.
\newblock {\em Dengue: guidelines for diagnosis, treatment, prevention and control}.
\newblock World Health Organization, 2nd edition, 2009.

\bibitem{WHO}
{WHO}.
\newblock {\em Vector-borne diseases: Dengue}.
\newblock Fact sheet No.~387, March 2014.
\newblock \url{http://www.who.int/mediacentre/factsheets/fs387/en/index2.html}

\bibitem{Wilder-Smith}
A. Wilder-Smith \emph{et al.}
\newblock The 2012 dengue outbreak in Madeira: exploring the origins.
\newblock {\em Euro Surveill.}, 19(8):pii=20718, 2014.

\end{thebibliography}
\end{document}